\documentclass[prb,amsfonts,amssymb,floats,twocolumn,superscriptaddress,aps]{revtex4}

\usepackage[final,dvips]{epsfig}
\usepackage{amsmath}
\usepackage{float}
\usepackage[caption = false]{subfig}
\usepackage{feynmp}
\usepackage{graphicx}

\usepackage{color}
\newcommand{\w}{\omega}
\newcommand{\W}{\Omega}
\newcommand{\wk}{\omega_{\vec k}}
\newcommand{\Wk}{\Omega_{\vec k}}
\newcommand{\gk}{\gamma_{\vec k}}

\newcommand{\ii}{\imath}   

\newcommand{\qq}{q}       
\newcommand{\qqc}{q_c}    
\newcommand{\lm}{\kappa} 

\newcommand{\x}{k}      

\newcommand{\vqa}{\Phi_{41}}  
\newcommand{\vqb}{\Phi_{42}}  
\newcommand{\vqc}{\Phi_{43}}  
\newcommand{\vqd}{\Phi_{44}}  

\newcommand{\vca}{\Phi_{31}}  
\newcommand{\vcb}{\Phi_{32}}  
\newcommand{\vcc}{\Phi_{33}}  
\newcommand{\vcd}{\Phi_{34}}  

\begin{document}

\unitlength=1mm

\title[]
{
Non-linear bond-operator theory and $1/d$ expansion for coupled-dimer magnets I: \\
Paramagnetic phase
}
\author{Darshan G. Joshi}
\affiliation{Institut f\"ur Theoretische Physik,
Technische Universit\"at Dresden, 01062 Dresden, Germany}
\author{Kris Coester}
\author{Kai P. Schmidt}
\affiliation{Theoretische Physik, Technische Universit\"at Dortmund,
Otto-Hahn-Str. 4, 44221 Dortmund, Germany}
\author{Matthias Vojta}
\affiliation{Institut f\"ur Theoretische Physik,
Technische Universit\"at Dresden, 01062 Dresden, Germany}

\date{\today}

\begin{abstract}
For coupled-dimer Heisenberg magnets, a paradigm of magnetic quantum phase transitions,
we develop a systematic expansion in $1/d$, the inverse number of space dimensions. The
expansion employs a formulation of the bond-operator technique and is based on the
observation that a suitably chosen product-state wavefunction yields exact
zero-temperature expectation values of local observables in the $d\to\infty$ limit, with
corrections vanishing as $1/d$. We demonstrate the approach for a model of dimers on a
hypercubic lattice, which generalizes the square-lattice bilayer Heisenberg model to
arbitrary $d$.
In this paper, we use the $1/d$ expansion to calculate static and dynamic observables at zero temperature in the paramagnetic singlet phase, up to the quantum phase transition, and compare the results with numerical data available for $d=2$.
Contact is also made with previously proposed refinements of bond-operator theory as well as with a perturbative expansion in the inter-dimer coupling.
In a companion paper, the present $1/d$ expansion will be extended to the ordered phase,
where it is shown to consistently describe the entire phase diagram
including the quantum critical point.
\end{abstract}


\maketitle


\section{Introduction}
\label{sec:intro}

Paramagnetic phases of quantum spin systems and their instabilities via quantum phase transitions (QPT)
have attracted enormous interest over the past two decades.\cite{ssbook,ss_natph,gia08,hvl}
Theoretical approaches can be roughly grouped into
(i) effective low-energy field theories, often combined with a renormalization-group treatment,
(ii) approximate microscopic calculations, e.g., using series expansions or auxiliary-particle approaches,
(iii) exact numerical methods, e.g., exact diagonalization or quantum Monte Carlo (QMC).
While coarse-grained field-theoretic techniques are well suited to capture universal
properties near criticality, a more quantitative connection to experiments and materials
often requires microscopic modelling. Here, a major problem on the analytical side is
that most approaches either contain uncontrolled approximations or are restricted to
describing a single phase while failing in crossing a QPT.
In this paper, we present a novel expansion method which does not suffer from these
restrictions.

We concentrate on an important class of systems with magnetic QPT, namely coupled-dimer
Heisenberg magnets\cite{ssbook,ss_natph,gia08} in space dimensions $d\geq 2$. In these
systems, realized in materials like TlCuCl$_3$, BaCuSi$_2$O$_6$, and Ba$_3$Cr$_2$O$_8$,
quantum spins form natural pairs (dimers) with typically strong antiferromagnetic
pairwise coupling, connected by a network of weaker inter-dimer couplings.
Such materials may display both paramagnetic and antiferromagnetic ground states, with
the QPT being accessible by varying pressure or magnetic field.

For coupled-dimer Heisenberg models of individual spins 1/2, bond operators were proposed
as an efficient auxiliary-particle description.\cite{bondop} In the original
formulation, four bond operators were introduced to describe the four states of the
Hilbert space of each dimer and combined with a mean-field approximation, yielding a
simple (but uncontrolled) description of the excitations of the paramagnetic phase in
terms of independent bosonic spin-1 particles (so-called triplons\cite{schmi03}).
Later, generalized bond operators were used for cases with larger Hilbert space per unit cell,
i.e., dimerized systems with spins $S>1/2$ or tetramerized systems.\cite{zhito96,kumar10,parame11,doretto14}
In addition, the bond-operator technique was generalized to magnetically ordered phases using a
suitable basis rotation in the Hilbert space of an isolated dimer\cite{sommer,penc} --
this enabled calculations across the entire phase diagram.
However, the description was mainly restricted to Gaussian fluctuations around a
saddle point, i.e., excitations were treated as non-interacting bosons, and a small
parameter controlling this approximation was not known.
Refined versions of the bond-operator technique have been developed to include
interactions between the triplons,\cite{chub95,kotov,alter} but their applicability
appears limited, again because of the lack of a systematic control parameter.

In this paper, we develop a systematic expansion in $1/d$ for coupled-dimer magnets.
Formally, this expansion is based on bosonic bond operators combined with suitable
projection operators to impose the required Hilbert-space constraint.
We show how to calculate thermodynamic and spectral properties order by order in $1/d$.
The use of $1/d$ as a physical small parameter ensures internal consistency:
As will be shown in a companion paper,\cite{ii} the expansions for the paramagnetic and
antiferromagnetic phases merge smoothly at the quantum phase transition which is obtained
as a continuous transition with a vanishing excitation gap. For a model with SU(2) spin
symmetry, the transverse spin excitations of the ordered phase are gapless at every order
in $1/d$, as required by Goldstone's theorem.

Although our approach is inspired by non-linear spin-wave theory, the most important
difference is that we work directly with quantum spins $S=1/2$, such that no semiclassical
approximation is possible. Instead, we expand about a dimer product state which is a
suitable reference state in the $d\to\infty$ limit. Technically, our expansion also
differs from spin-wave theory: the latter features explicit factors of $1/S$ in the
Hamiltonian, whereas in our approach factors of $1/d$ are only {\em generated} via
momentum summations.


\subsection{$1/d$ expansion and quantum criticality}
\label{sec:mfchar}

Before diving into details, we discuss the general question how a systematic $1/d$ expansion can access quantum critical behavior.
First, it is important to realize that any Taylor expansion for an observable assumes analyticity (as function of both $1/d$ and other control parameters), i.e., is {\em a priori} not compatible with singular behavior found at quantum critical points.
Second, we recall that the critical exponents of the magnets under consideration are locked to mean-field values\cite{goldenfeld} above the upper-critical dimension (here $d_c^+=3$).\cite{ssbook} Hence, critical exponents must take mean-field values to all orders in a $1/d$ expansion, and non-trivial exponents are not accessible.

Interestingly, we can use the mean-field nature of the transition to our advantage, namely by identifying observables which are {\em analytic} even at criticality. For instance, the excitation gap $\Delta$ of the disordered phase varies with the distance $t$ to the critical point as $\Delta \propto t^{\nu z}$, with the correlation length and dynamical exponents $\nu=1/2$, $z=1$. This implies that $\Delta^2 \propto t$ is analytic and hence amendable to a $1/d$ expansion. Similar considerations apply to the order parameter (in the ordered phase) and other observables and will be used throughout the paper to extract critical properties.

Notwithstanding, there are physics aspects which cannot be captured by a $1/d$ expansion, for instance the zero-temperature damping of excitations. As will be shown in the paper, the inverse lifetimes induced by interactions are exponentially suppressed at large $d$, and hence there is no damping to all orders in $1/d$.


\subsection{Model and large-$d$ limit}

\begin{figure}[!t]
\includegraphics[width=0.48\textwidth]{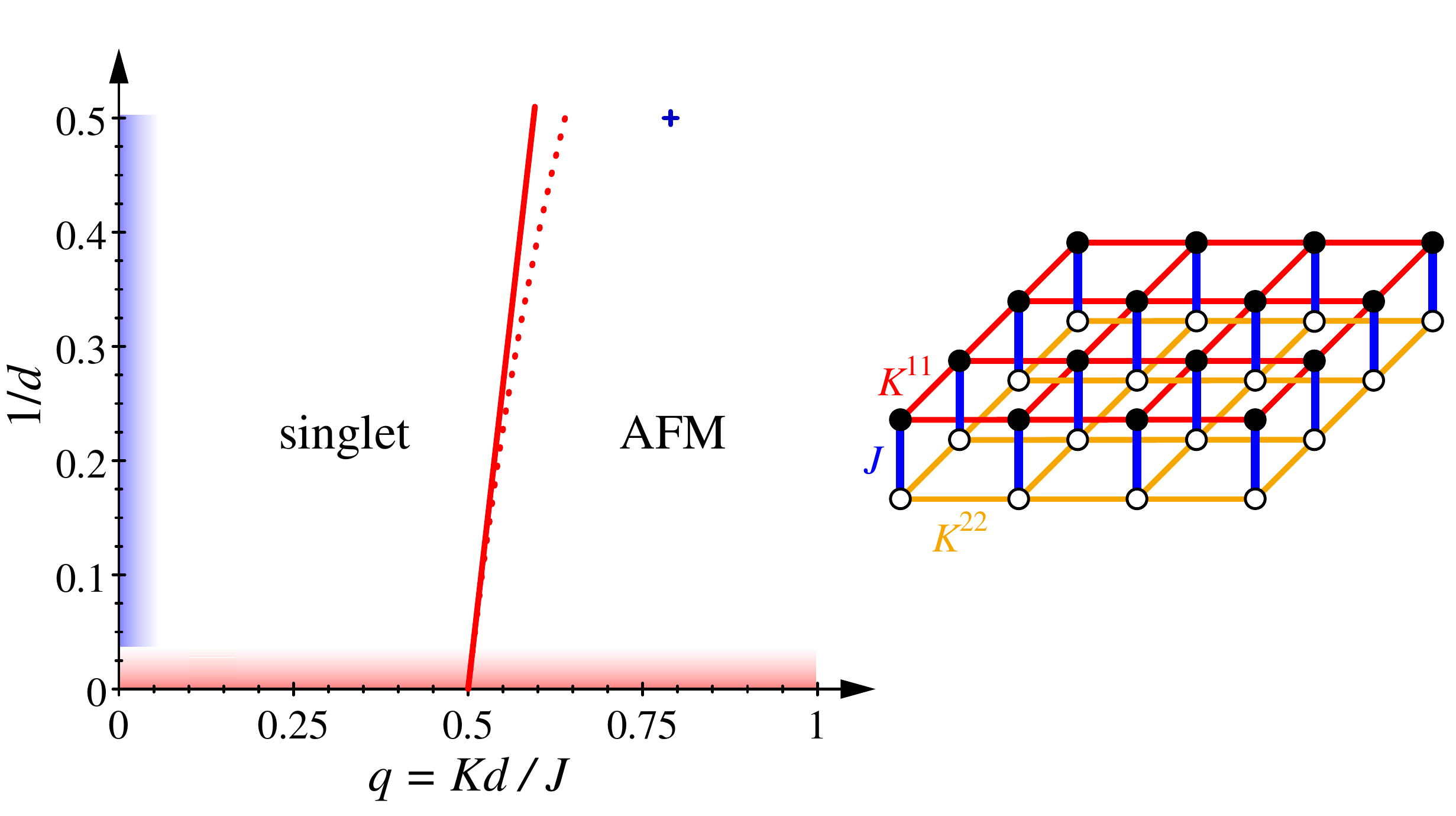}
\caption{
Left: Phase diagram of the coupled-dimer model \eqref{hh} on the hypercubic lattice as function of the
control parameter $\qq = Kd/J$ and the inverse spatial dimension $1/d$. A QPT separates the
paramagnetic singlet phase from the AFM phase. The solid line shows our result \eqref{qqcexp} for the phase boundary $\qqc$ to order $1/d$ for the symmetric case $\lm=0$ \eqref{lamdef}; the dashed line represents the solution of the equation $\Delta(\qq)=0$ with $\Delta(\qq)$ from Eq.~\eqref{gap-s}.
%
The cross marks the numerically exact result for $d=2$ obtained in
Ref.~\onlinecite{sandvik06}.
The shaded areas indicate the points of departure for the large-$d$ expansion (red) and the small-$K/J$
expansion (blue, Section~\ref{sec:smallk}), respectively.
Right: Sketch of the model in $d=2$, with solid (hollow) dots denoting the spins 1 (2) of each dimer.
}
\label{fig:pd1}
\end{figure}

The $1/d$ expansion is applicable to coupled-dimer Heisenberg magnets of spin $1/2$, with
the general Hamiltonian
\begin{equation}
\label{h}
\mathcal{H} =
\sum_i J_i \vec{S}_{i1} \cdot \vec{S}_{i2} +
\sum_{ii'mm'} K_{ii'}^{mm'} \vec{S}_{im} \cdot \vec{S}_{i'm'}
\end{equation}
where the indices $i,i'$ refer to sites on a regular lattice of dimers, and $m,m'=1,2$
refer to the individual spins on each dimer.
For most of the paper, we will be specifically concerned with dimers on a hypercubic
lattice in space dimension $d$, Fig.~\ref{fig:pd1}, where
\begin{equation}
\label{hh}
\mathcal{H} =
J \sum_i \vec{S}_{i1} \cdot \vec{S}_{i2} +
\sum_{\langle ii'\rangle} (K^{11} \vec{S}_{i1} \cdot \vec{S}_{i'1} + K^{22} \vec{S}_{i2} \cdot \vec{S}_{i'2})
\end{equation}
and $\sum_{\langle ii'\rangle}$ now denotes a summation over pairs of nearest-neighbor dimer
sites on the hypercubic lattice. We have allowed for different couplings within the hypercubic lattices corresponding to $m=1,2$, and define
\begin{equation}
K = \frac{K^{11} + K^{22}}{2}\,,~~\lm K = \frac{K^{11} - K^{22}}{2}\,,
\label{lamdef}
\end{equation}
where $\lm$ is an asymmetry parameter.
For $d=1$ and 2 the spin lattice of $\mathcal{H}$ in Eq.~\eqref{hh} corresponds to the much-studied two-leg ladder and square-lattice bilayer magnets, respectively.

A non-trivial limit $d\to\infty$ is obtained if the inter-dimer coupling constant $K$ is scaled as $1/d$ in order to preserve a non-trivial competition between the $K$ and $J$ terms in the Hamiltonian \eqref{hh}.\cite{scalenote} Hence, for $d\geq 2$ and $K,J>0$, the dimensionless parameter
\begin{equation}
\qq = \frac{Kd}{J}
\end{equation}
controls a quantum phase transition between a singlet paramagnet at small $\qq$ and
an antiferromagnet with ordering wavevector $(\pi,\pi,\ldots)$ at large $\qq$.
For $d=2$ this transition occurs at\cite{sandvik06} $\qqc = 0.793$ for $\lm=0$ 
and $\qqc = 0.720$ for $|\lm|=1$. 

A suitable starting point for an expansion is a product wavefunction $|\psi_0\rangle =
\prod_i |\psi\rangle_i$ where $|\psi\rangle_i$ denotes an arbitrary normalized state of
dimer $i$. A simple variation of $\langle\psi_0|\mathcal{H}|\psi_0\rangle$ with
$\mathcal{H}$ from Eq.~\eqref{hh} yields a transition at $\qqc = 1/2$;
for $\qq<\qqc$ the variational minimum is of course found for the singlet, $|\psi\rangle_i =
(|\uparrow\downarrow\rangle_i - |\downarrow\uparrow\rangle_i)/\sqrt{2}$, while a linear
combination of singlet and one triplet minimizes $\langle\psi_0|\mathcal{H}|\psi_0\rangle$ for
$\qq>\qqc$. Gaussian fluctuations around this product state have been analyzed
previously.\cite{sommer}

Here we show that the product state $|\psi_0\rangle$ delivers exact expectation values for local observables in the limit \mbox{$d\to\infty$} for any $\qq$, i.e., corrections from non-local fluctuations vanish in this limit. The reason is that fluctuation effects tend to average out in the limit of large connectivity. This then paves the way for a systematic expansion in $1/d$, described in the body of the paper.

Note that this does {\em not} imply that $|\psi_0\rangle$ becomes the exact ground state
as $d\to\infty$; as we show below, corrections to the wavefunction are generally non-vanishing in
this limit. This also distinguishes our limit\cite{scalenote} $d\to\infty$ at fixed $\qq$ (i.e.~$K/J\propto
1/d$) from the limit of weak inter-dimer coupling, $K/J \to 0$ at fixed $d$; in the latter, a singlet product state is trivially the exact ground state.

\subsection{Summary of results}

We now quote our main results of the $1/d$ expansion applied to the model \eqref{hh} for $\lm=0$; results for the asymmetric case of nonzero $\lm$, together with an extensive discussion, can be found in the body of the paper.

The quantum critical point, Fig.~\ref{fig:pd1}, is located at:
\begin{equation}
\label{qqcexp}
\qqc = \frac{1}{2} + \frac{3}{16} \frac{1}{d} + \mathcal{O}\Big(\frac{1}{d^2}\Big)\,.
\end{equation}
In the paramagnetic phase, the triplet mode dispersion near the ordering wavevector $\vec{Q}$ can be parameterized by
\begin{equation}
\label{quadpara}
\Wk^2 = \Delta^2 + \frac{c^2}{d} (\vec{k}-\vec{Q})^2 \,.
\end{equation}
The energy gap $\Delta$ behaves as
\begin{equation}
\frac{\Delta^{2}}{J^2} = 1-2\qq + \frac{1}{d} (2 \qq^{2} -  \qq^{3})
+\mathcal{O}\left(\frac{1}{d^2}\right)\,,
\label{gap-s}
\end{equation}
it closes at $\qqc$ according to
\begin{equation}
\label{deltanearqc}
\frac{\Delta}{J} = \left[\sqrt{2} - \frac{5}{8 \sqrt{2} d} +
\mathcal{O}\Big(\frac{1}{d^2}\Big) \right]\sqrt{\qqc - \qq},
\end{equation}
corresponding to mean-field exponents $\nu=1/2$ and $z=1$,
and the critical velocity has the following $1/d$ expansion:
\begin{equation}
\label{cnearqc}
\frac{c}{J} = \frac{1}{\sqrt{2}} + \frac{5}{16\sqrt{2} d} + \mathcal{O}\Big(\frac{1}{d^2}\Big).
\end{equation}

Corresponding results for the antiferromagnetically ordered phase will be given in the
companion paper Ref.~\onlinecite{ii}.


\subsection{Relation to previous work}

A number of approaches have been used in the literature to treat triplet excitations in
coupled-dimer magnets beyond the limit of non-interacting bosons.

A first approach is to systematically expand in $K/J$ which can be done in principle up
to high orders.\cite{gelfand90,hida92,oitmaa96,gsu00} This naturally yields accurate
results for small $K/J$, but cannot reliably cover the regime close to the critical point
and beyond. We will show that our $1/d$ expansion, when applied for small $\qq$, delivers
results consistent with an expansion in $K/J$ if the latter is done for large $d$.

A second approach is to include interactions in an approximate fashion into the
bond-operator theory. An efficient treatment of the hard-core repulsion of triplet
excitations has been proposed by Kotov {\em et al.}\cite{kotov} via Brueckner theory;
this yields an accurate value for the location of the critical point for the bilayer
Heisenberg model. Brueckner theory is designed to work in the limit of small triplet
density, and we will compare its results with our systematic results for large $d$.
We note that attempts to generalize the Brueckner approach to the ordered phase lead to
either a violation of Goldstone's theorem or to the QPT being (erroneously) rendered
first order.\cite{kotov_ordered} These problems appear to be rooted in the lack of a
systematic expansion parameter controlling the approximation.
An earlier treatment by Chubukov and Morr,\cite{chub95} inspired by non-linear spin-wave
theory, works in both phases, but suffers from divergencies at higher orders, probably
because it lacks a small control parameter as well.
Recently, Collins {\em et al.}\cite{alter,alter2} proposed to implement the hard-core
repulsion of triplet excitations using projection operators and used this to calculate
properties in the paramagnetic phase in the spirit of a small-$K$ expansion.

A third approach, due to Jensen,\cite{jensen11} is based on a $1/z$ expansion for Green's
functions, where $z$ is the lattice coordination number. Similar to earlier
work,\cite{jensen84,jensen94} it has been used to calculate excitation energies in the
paramagnetic phase, but a systematic analysis order by order in $1/z$ has not been
performed to our knowledge. Related $1/z$ expansions have been applied to other types
of lattice models,\cite{stinchcombe73,bak75,fishman90} but the vicinity of a QPT has not been studied.

It is worth mentioning that various methods exist to describe spin excitations of
semiclassically ordered states beyond the limit of non-interacting bosons. The most
prominent microscopic approach is non-linear spin-wave theory, and we will make contact
between this and our method in Ref.~\onlinecite{ii}. Here we only point out that these
semiclassical methods cannot cover the regime near quantum criticality of the
models (\ref{h},\ref{hh}), mainly because longitudinal fluctuations are neglected.

Finally, a certain class of infinite-range Heisenberg models was investigated in Ref.~\onlinecite{mudry89} where valence-bond states could be stabilized via suitable perturbations. In both the infinite-range case and our $d\to\infty$ limit the number of interaction partners of each spin diverges.


\subsection{Outline}

The body of the paper is organized as follows:
In Section~\ref{sec:bond} we introduce the formulation of the bond-operator
representation to be employed in the paper.
This is used in Section~\ref{sec:ham} to construct an exact interacting Hamiltonian of
triplet excitations on top of a singlet background.
In Section~\ref{sec:expa} we develop the $1/d$ expansion in the paramagnetic phase. Starting
from the singlet-product-state description, we first show that fluctuation corrections to
thermodynamic quantities vanish as $d\to\infty$. We then demonstrate how to evaluate
those corrections, as well as corrections to the triplon dispersion, in a power series in
$1/d$ and present explicit results for the hypercubic dimer model \eqref{hh}. Particular attention is paid to the asymmetric case, $\lm\neq0$ \eqref{lamdef}, as this induces cubic triplon vertices which are absent in the symmetric situation.
Section~\ref{sec:smallk} provides an important cross-check for our approach: We calculate
observables in an expansion in $K/J$, i.e., the relative strength of the inter-dimer coupling, for
the hypercubic dimer model \eqref{hh} in arbitrary dimension $d$. The results of this and
our $1/d$ expansion are shown to be consistent in the combined limit of small $1/d$ and
small $\qq$.
In Section~\ref{sec:olatt} we discuss aspects of our method beyond the thermodynamic-limit hypercubic-lattice case, including large-$d$ generalizations of given finite-$d$ lattice models.
The concluding Section~\ref{sec:concl} describes possible extensions and further applications of our
method.
Technical details are relegated to various appendices.

A companion paper, Ref.~\onlinecite{ii}, will be devoted to the extension of the $1/d$
expansion to magnetically ordered phases of coupled-dimer models.


\section{Bond operators and projection}
\label{sec:bond}

Bond-operator theory employs a slave-particle description of the states of each dimer
$i$. We denote those states by $|t_k\rangle_i$,
$k=0, \ldots, 3$, where
$|t_0\rangle=
(|\uparrow\downarrow\rangle -|\downarrow\uparrow\rangle)/\sqrt{2}$ is the spin-0 singlet
state, and
$|t_1\rangle =
(-|\uparrow\uparrow\rangle+|\downarrow\downarrow\rangle)/\sqrt{2}$,
$|t_2\rangle =
\ii(|\uparrow\uparrow\rangle+|\downarrow\downarrow\rangle)/\sqrt{2}$,
$|t_3\rangle =
(|\uparrow\downarrow\rangle+|\downarrow\uparrow\rangle)/\sqrt{2}$
correspond to the spin-1 triplet, and $\ii$ is the imaginary unit.

The initial bond-operator approach of Sachdev and Bhatt\cite{bondop} introduced four
bosonic operators $t_{ik}^\dagger$ which create these states out of a fictitious vacuum,
$|t_k\rangle_i = t_{ik}^\dagger |vac\rangle_i$, leading to the following representation of the
original spin operators in terms of bond bosons:
\begin{equation}
{S}_{i1,2}^\alpha = \frac{1}{2}\left(\pm t^\dagger_{i\alpha} t_{i0} \pm t_{i0}^\dagger t_{i\alpha}
          - \ii\epsilon_{\alpha\beta\gamma}t^\dagger_{i\beta} t_{i\gamma} \right),
\label{spin-bondop}
\end{equation}
where $\alpha=1,2,3\equiv x,y,z$, and the upper (lower) sign corresponds to spin 1 (2) of each dimer.
The constraint
\begin{equation}
\sum_{k=0}^3  t^\dagger_{ik} t_{ik} =1
\end{equation}
then defines the physical Hilbert space.
For the subsequent treatment the singlet operator was condensed, $t_{i0} \rightarrow
\langle t_{i0}\rangle = s$, and the constraint was treated in a mean-field fashion via a
Lagrange multiplier $\mu$, such that $\sum_{\alpha=1}^3 \langle t^\dagger_{i\alpha}
t_{i\alpha} \rangle + s^2 =1$. In the Hamiltonian, only bilinear terms in the $t_\alpha$
operators were kept, amounting to a harmonic approximation for the triplet excitations,
and the mean-field parameters $s$ and $\mu$ were determined variationally.

\subsection{Excitations as hard-core bosons}

An alternative approach is due to Kotov {\em et al.}\cite{kotov} It starts by
reformulating the Hilbert space in terms of a singlet vacuum and triplet particles.
Then, the operators $t_{i\alpha}^\dagger$ create excitations on top of the singlet
background state (i.e.~represent the $t_{i\alpha}^\dagger t_{i0}$ operator of Sachdev and
Bhatt\cite{bondop}), and singlet operators no longer appear.
The triplet excitations obey the hard-core constraint
\begin{equation}
\label{hardcore}
\sum_{\alpha=1}^3  t^\dagger_{i\alpha} t_{i\alpha} \leq 1.
\end{equation}
A harmonic approximation to the resulting Hamiltonian, also ignoring the hard-core
constraint, is similar in spirit to linear spin-wave theory and has been employed in a
number of papers.\cite{sommer,penc,brenig97,MatsumotoNormandPRL,MVTU}

Ref.~\onlinecite{kotov} proposed to go beyond the harmonic approximation by encoding the
hard-core constraint as an infinite on-site repulsion,
\begin{equation}
\label{HU}
\mathcal{H}_U = U\sum_{i\alpha\beta}
t^\dagger_{i\alpha}t^\dagger_{i\beta}t_{i\alpha}t_{i\beta}, \quad
U\rightarrow\infty\;,
\end{equation}
and treating this via the so-called Brueckner approach which involves a self-consistent
summation of ladder diagrams and is controlled in the small-density limit. In addition,
quartic triplet terms were included in a Hartree-Fock approximation. In
Appendix~\ref{app:kotov} we will discuss the possibility to generate a $1/d$ expansion
using this approach.

\subsection{Projection operators}
\label{sec:proj}

More recently, Collins {\em et al.} \cite{alter} implemented the hard-core constraint
\eqref{hardcore} for the $t_{i\alpha}$ using projection operators which suppress any matrix
element of observables between states inside and outside the physical Hilbert space.
Using such projection operators, the spin operators ${S}_{im}^\alpha$ acquire the
following representation in terms of the triplet operators $t_{i\alpha}$:
\begin{equation}
\label{sproj}
{S}_{i1,2}^\alpha = \frac{1}{2} \left(
\pm t_{i\alpha}^\dagger P_i
\pm P_i t_{i\alpha}
- \ii \epsilon_{\alpha\beta\gamma} t_{i\beta}^\dagger t_{i\gamma} \right)
\end{equation}
where $P_i$ prevents the creation of more than one triplet excitation on site $i$.

In our calculations we shall adopt this procedure and, as in Ref.~\onlinecite{alter}, we
will use the projector
\begin{equation}
\label{linproj}
P_i = 1- \sum_\gamma t_{i\gamma}^\dagger t_{i\gamma}.
\end{equation}
It can then be shown that the $\vec{S}_{im}$ \eqref{sproj} obey the standard spin commutation relations inside the physical Hilbert space. Other choices of the projector are not advantageous, as explained in Appendix~\ref{app:commu}.


\section{Hamiltonian and perturbation theory}
\label{sec:ham}

In this section we discuss how to generate the perturbative expansion in $1/d$ for
coupled-dimer models. We will primarily deal with the hypercubic-lattice case as in
Eq.~\eqref{hh}; differences arising from other lattice geometries will be discussed in
Section~\ref{sec:olatt}.

\subsection{Real-space bond-operator Hamiltonian}

Using the representation \eqref{sproj} of spin operators, the Hamiltonian \eqref{hh}
takes the following form
\begin{align}
\mathcal{H} &=
J \sum_{i\alpha} (t^\dagger_{i\alpha} t_{i\alpha} - \frac{3}{4}) \notag\\
&+ \frac{K}{2} \sum_{\langle ii'\rangle\alpha}
(t^\dagger_{i\alpha} P_i P_{i'} t_{i'\alpha} + t^\dagger_{i\alpha} P_i t^\dagger_{i'\alpha} P_{i'} + h.c.) \notag\\
&- \frac{\lm K}{2} \sum_{\langle ii'\rangle\alpha\beta\gamma} \!
\epsilon_{\alpha\beta\gamma}
\left[(\ii t^\dagger_{i\alpha} P_i t^\dagger_{i'\beta} t_{i'\gamma} + h.c.) + (i \leftrightarrow i')\right]\notag\\
&+ \frac{K}{2} \sum_{\langle ii'\rangle\alpha\beta}
(t^\dagger_{i\alpha}t^\dagger_{i'\beta}t_{i\beta}t_{i'\alpha} -
 t^\dagger_{i\alpha}t^\dagger_{i'\alpha}t_{i\beta}t_{i'\beta}) \,.
\label{hhp}
\end{align}

Inserting the projector \eqref{linproj} into Eq.~\eqref{hhp} results in a Hamiltonian
with non-linear couplings up to 6th order,
\begin{equation}
\label{hhexp}
\mathcal{H} = \mathcal{H}_0 + \mathcal{H}_2 + \mathcal{H}_3 + \mathcal{H}_4 + \mathcal{H}_5 + \mathcal{H}_6,
\end{equation}
where $\mathcal{H}_n$ contains $n$ triplet operators.
$\mathcal{H}_0 = -\frac{3}{4} J N$ is the energy of the product state $|\psi_0\rangle$,
with $N$ the number of dimer sites. The remaining even-$n$ terms read:
\begin{align}
\label{hh2}
\mathcal{H}_2 &=
J \sum_{i,\alpha} t^\dagger_{i\alpha} t_{i\alpha}
+ \frac{K}{2} \sum_{\langle ii'\rangle\alpha} (t^\dagger_{i\alpha} t_{i'\alpha} + t^\dagger_{i\alpha} t^\dagger_{i'\alpha} + h.c.)\,,
\\
%
\mathcal{H}_4 &=
\frac{K}{2} \sum_{\langle ii'\rangle\alpha\beta}
(t^\dagger_{i\alpha}t^\dagger_{i'\beta}t_{i\beta}t_{i'\alpha} -
t^\dagger_{i\alpha}t^\dagger_{i'\alpha}t_{i\beta}t_{i'\beta}) \notag\\
&-
\frac{K}{2} \sum_{\langle ii'\rangle\alpha\beta}
 (t^\dagger_{i\alpha} t^\dagger_{i\beta} t_{i\beta} t_{i'\alpha} + t^\dagger_{i\alpha} t^\dagger_{i'\beta} t_{i'\beta} t_{i'\alpha} \notag\\
&~~~~~+ t^\dagger_{i\alpha} t^\dagger_{i'\alpha} t^\dagger_{i\beta} t_{i\beta}
+ t^\dagger_{i\alpha} t^\dagger_{i'\alpha} t^\dagger_{i'\beta} t_{i'\beta} + h.c.)\,,
\label{hh4}
\end{align}
and
\begin{align}
\mathcal{H}_6 &=
\frac{K}{2} \sum_{\langle ii'\rangle} \sum_{\alpha, \beta, \gamma}
\left( t^\dagger_{i\alpha} t^\dagger_{i \beta} t^\dagger_{i'\alpha} t^\dagger_{i'\gamma} t_{i \beta}  t_{i' \gamma} \right. \nonumber \\
&~~~~~~~~~~~~~~~~+ t^\dagger_{i\alpha} t^\dagger_{i \beta} t^\dagger_{i' \gamma} t_{i \beta} t_{i' \gamma} t_{i'\alpha} + h.c.
\left. \right) \,.
\label{hh6}
\end{align}
For asymmetric couplings, $K^{11}\neq K^{22}$, the following odd-$n$ terms occur in addition:
\begin{align}
\mathcal{H}_3 &= - \frac{\lm K}{2} \sum_{\langle ii'\rangle\alpha\beta\gamma} \!
\epsilon_{\alpha\beta\gamma}
\left[(\ii t^\dagger_{i\alpha} t^\dagger_{i'\beta} t_{i'\gamma} + h.c.) + (i \leftrightarrow i')\right]
\label{hh3}
\end{align}
and
\begin{align}
\mathcal{H}_5 &=  \frac{\lm K}{2} \!\!\! \sum_{\langle ii'\rangle\alpha\beta\gamma\kappa} \!\!\!
\epsilon_{\alpha\beta\gamma}
\left[(\ii t^\dagger_{i\alpha} t^\dagger_{i\kappa} t^\dagger_{i'\beta} t_{i'\gamma} t_{i\kappa} + h.c.) + (i \leftrightarrow i')\right].
\label{hh5}
\end{align}
Cubic terms of the form \eqref{hh3} have been discussed in the context of two-particle decay of triplet excitations at elevated energies both experimentally\cite{stone,masuda} and theoretically.\cite{kolezhuk06,zhito06}

Two remarks concerning the full Hamiltonian are in order: First, the strength of all non-linear coupling is set by $K$. Second, individual pieces of $\mathcal{H}$ violate the constraint \eqref{hardcore}, and only an infinite-order treatment will restore the constraint exactly. In the expansion described below, the constraint is expected to be obeyed order by order in $1/d$.

\subsection{Bilinear part}

The free-triplon part \eqref{hh2} of the Hamiltonian takes the following form in Fourier
space:
\begin{equation}
\label{hh2k}
\mathcal{H}_2 = \sum_{\vec k\alpha} \left[
A_{\vec k} t^\dagger_{\vec k\alpha} t_{\vec k\alpha} +
\frac{B_{\vec k}}{2} (t^\dagger_{\vec k\alpha}t^\dagger_{-\vec k\alpha} + h.c.)
\right]
\end{equation}
where
\begin{equation}
\label{bareAB}
A_{\vec k} = J + B_{\vec k}\,,~
B_{\vec k} = \qq J \gk
\end{equation}
and the structure factor of the interaction
\begin{equation}
\gk = \frac{1}{d} \sum_{n=1}^d \cos k_n
\label{gammadef}
\end{equation}
which is normalized such that $-1\leq\gk\leq1$.
The bilinear Hamiltonian \eqref{hh2k} is solved by a standard Bogoliubov transformation,
\begin{equation}
\label{bogol}
t_{\vec{k}\alpha} =
u_{\vec{k}}\tau_{\vec{k}\alpha}+v_{\vec{k}}\tau^\dagger_{-\vec{k}\alpha},
\end{equation}
which transforms it into
\begin{equation}
\label{hh2kd}
\mathcal{H}_2 = \sum_{\vec k\alpha} \wk \tau^\dagger_{\vec k\alpha} \tau_{\vec k\alpha}
+ \frac{3}{2} \sum_{\vec k} (\wk - A_{\vec{k}})
\end{equation}
with mode energies
\begin{equation}
\label{om0}
\wk  = \sqrt{A_{\vec{k}}^2-B_{\vec{k}}^2}
= J \sqrt{1+2\gk\qq}
\end{equation}
and Bogoliubov coefficients
\begin{equation}
\label{eq3}
u_{\vec{k}}^2, v_{\vec{k}}^2 = \pm \frac{1}{2} +
\frac{A_{\vec{k}}}{2\wk}\,,~~
u_{\vec k} v_{\vec k} = - \frac{B_{\vec k}}{2\wk}.
\end{equation}

\subsection{Large-$d$ limit and perturbation theory}

The physics of the bilinear Hamiltonian $\mathcal{H}_2$, usually referred to as harmonic
approximation, can be used to discuss the limit of large dimensions $d$. Due to the
anomalous piece, pairs of triplets get admixed into the ground state. The
wavefunction in harmonic approximation can be written as
\begin{equation}
|\psi\rangle \propto \exp\left(
\sum_{\vec k\alpha} \frac{v_{\vec k}}{u_{\vec k}} \, t^\dagger_{\vec k\alpha}t^\dagger_{-\vec k\alpha}
\right) |\psi_0\rangle\,.
\end{equation}
The {\em local} triplet density evaluates to
\begin{equation}
\langle\psi| t_{i\gamma}^\dagger t_{i\gamma} |\psi\rangle = \frac{1}{N} \sum_{\vec k} v_{\vec k}^2
~\overset{d\to\infty}{=}~ \frac{\qq^2}{8d}\,,
\end{equation}
see Appendix~\ref{app:expval}. Similarly, expectation values like $\langle
t_{i\gamma} t_{j\gamma} \rangle$, with $i,j$ being neighboring sites, vanish as \mbox{$d\to\infty$}.
This implies, as announced, that the product state $|\psi_0\rangle$ yields exact
ground-state expectation values in the limit $d\to\infty$.
All corrections can be systematically evaluated in power series in $1/d$ -- this is
the subject of this paper.

Technically, we shall calculate observables for the model \eqref{hh} by an expansion in
the non-linear couplings $\mathcal{H}_{3,4,5,6}$ in \eqref{hhexp} using standard
diagrammatic perturbation theory. While there is no small parameter controlling such an
expansion in arbitrary fixed $d$, it will become clear that, for large $d$, perturbative contributions
to observables are suppressed by an increasing number of powers of $1/d$ with increasing
order in perturbation theory.

The origin of this suppression lies in the momentum summations for large $d$ which involve powers of the interaction structure factor $\gk$ \eqref{gammadef}: For a {\em typical} $\vec k$, $\gk$ is a sum of $d$ ``random'' numbers which tend to average out, such that the magnitude of $\gk$ for typical $\vec k$ scales as\cite{dmft} $1/\sqrt{d}$. In a momentum sum, most $\vec k$ are typical, such that $\gk$ can be used as a formal expansion parameter, see Appendix~\ref{app:expval}. The non-locality of the interactions in $\mathcal{H}$ then ensures that the perturbation theory can be truncated. However, the structure of the expansion is different from that of a loop expansion, i.e., diagrams with different numbers of loops contribute to any given order in $1/d$.

\subsection{Normal-ordered Hamiltonian}

Diagrammatic perturbation theory requires interactions terms which are normal-ordered in the $\tau_{\vec k\alpha}$, i.e., the operators which diagonalize the free-particle piece of $\mathcal{H}$. Upon expressing the non-linear couplings $\mathcal{H}_{4,6}$ in terms of the $\tau_{\vec k\alpha}$, normal ordering generates additional bilinear terms.
To deal with those, two different strategies have been employed in the spin-wave literature:
(i) A Bogoliubov transformation is used to diagonalize the leading-order bilinear terms,
i.e., the ones from $\mathcal{H}_2$, and the bilinear terms obtained from normal-ordering
of $\mathcal{H}_{4,6}$ are treated perturbatively.\cite{igarashi92,hamer92}
(ii) A Bogoliubov transformation is used to diagonalize {\em all} bilinear terms (up to the
order calculated) simultaneously; this then leads to a self-consistent equation for the
Bogoliubov coefficients.\cite{gochev94,alter}
For our $1/d$ expansion -- in particular at criticality and in the ordered phase\cite{ii} -- we found
it advantageous to employ strategy (i), because strategy (ii) would imply the necessity for a $1/d$
expansion of the $u_{\vec k}$, $v_{\vec k}$, and $\wk$ involved in the Bogoliubov transformation, which
is ill-defined if the leading-order $\w_{\vec k}$ vanishes. (As we show below, a $1/d$ expansion for
$\wk^2$ is well-defined instead.)

Hence, we employ the leading-order Bogoliubov transformation according to
Eqs.~\eqref{bogol}, \eqref{eq3}, and \eqref{bareAB}, to
generate a normal-ordered Hamiltonian in terms of the $\tau_{\vec k\alpha}$. This
Hamiltonian takes the form
\begin{equation}
\label{hpexp}
\mathcal{H} = \mathcal{H}'_0 + \mathcal{H}'_2 + \mathcal{H}'_3 + \mathcal{H}'_4 + \mathcal{H}'_5 + \mathcal{H}'_6
\end{equation}
where the $\mathcal{H}'_n$ now contain $n$ transformed $\tau$ operators and can be
obtained by a straightforward but tedious calculation.\cite{alter2}
Here we include terms up to 4th order in the $t$ operators -- this will be shown to be
sufficient to obtain the complete set of corrections to order $1/d$ to the mode
dispersion -- and use the explicit form of $A_{\vec k}$ and $B_{\vec k}$ in
Eq.~\eqref{bareAB}, assuming $u_{\vec k}=u^\ast_{\vec k}=u_{-\vec k}$ and $v_{\vec
k}=v^\ast_{\vec k}=v_{-\vec k}$.
The constant is
\begin{align}
\mathcal{H}'_0
&= 3JN \bigg[ -\frac{1}{4} + R_2 + \qq(R_3+R_4) \notag\\
& - 2\qq (R_1 + 4 R_2)(R_3 + R_4)  \notag \\
&-\frac{q}{N}\big[ \sum_{\vec{k}} u_{\vec{k}}v_{\vec{k}} R'_{3}(\vec{k})
  - \sum_{\vec{k}} v_{\vec{k}}^{2} R'_{4}(\vec{k})\big] \bigg]
\label{hp0}
\end{align}
which involves the abbreviations
\begin{align}
R_1 &= \frac{1}{N} \sum_{\vec k} u_{\vec k} v_{\vec k}\,,~~~~~R_2 = \frac{1}{N} \sum_{\vec k} v^2_{\vec k} \notag\\
R_3 &= \frac{1}{N} \sum_{\vec k} \gk u_{\vec k} v_{\vec k}\,, ~~R_4 = \frac{1}{N} \sum_{\vec k} \gamma_{\vec k} v^2_{\vec k}
\label{rdef}
\end{align}
and
\begin{align}
R'_3 (\vec{k'})&= \frac{1}{N} \sum_{\vec k} \gamma_{\vec{k'} - \vec k} u_{\vec k} v_{\vec k}\, \notag\\
R'_4 (\vec{k'})&= \frac{1}{N} \sum_{\vec k} \gamma_{\vec{k'} - \vec k} v^2_{\vec k}.
\label{rdef1}
\end{align}
As explicitly shown in Appendix~\ref{app:expval}, the $R_{1\ldots4}$ are suppressed in the large-$d$ limit at least as $1/d$ due to the properties of the large-$d$ momentum summations over $\gamma_{\vec k}$.

The bilinear $\tau$ terms can be split as $\mathcal{H}'_2 = \mathcal{H}'_{2a} +
\mathcal{H}'_{2b}$ where
\begin{equation}
\label{hp2a}
\mathcal{H}'_{2a} =
\sum_{\vec k\alpha} \wk \tau^\dagger_{\vec k\alpha} \tau_{\vec k\alpha}
\end{equation}
is the leading-order piece from $\mathcal{H}_2$, and
\begin{equation}
\label{hp2b}
\mathcal{H}'_{2b} =
\sum_{\vec k\alpha}
\left[
C_{\vec k} \tau^\dagger_{\vec k\alpha} \tau_{\vec k\alpha} +
\frac{D_{\vec k}}{2} (\tau^\dagger_{\vec k\alpha} \tau^\dagger_{-\vec k\alpha} + h.c.)
\right]
\end{equation}
contains the bilinear terms generated from normal-ordering of $\mathcal{H}_4$, with
\begin{widetext}
\begin{align}
\label{eq:ck}
C_{\vec k} &= \qq J \big[ 2(u_{\vec k}^{2} + v_{\vec k}^{2})R'_4 - 4 u_{\vec k} v_{\vec k} R'_3
-(2 \gk R_1 + 8 \gk R_2) (u_{\vec k} + v_{\vec k})^{2}
-4 (R_3 + R_4) (2 u_{\vec k}^{2} + 2 v_{\vec k}^{2} + u_{\vec k}v_{\vec k})
\big]\,, \\
D_{\vec k} &= \qq J \big[ 4u_{\vec k} v_{\vec k}R'_4 - 2 (u_{\vec k}^{2} + v_{\vec k}^{2}) R'_3
-(2 \gk R_1 + 8 \gk R_2) (u_{\vec k} + v_{\vec k})^{2}
-2 (R_3 + R_4) ( u_{\vec k}^{2} +  v_{\vec k}^{2} + 8 u_{\vec k}v_{\vec k})
\big] \,.
\label{eq:dk}
\end{align}
Given the behavior of $R_{1\ldots4}$ in the large-$d$ limit, all terms in both $C_{\vec k}$ and $D_{\vec k}$ are of order $1/d$ or smaller, such that the contribution of $\mathcal{H}'_{2b}$ is suppressed relative to $\mathcal{H}'_{2a}$ in this limit.
The quartic term is
\begin{align}
\label{hp4}
\mathcal{H}'_4 &= \frac{1}{N} \sum_{1234}
\big[\delta_{1+2+3+4} \vqa
(\tau^\dagger_{1\alpha}\tau^\dagger_{2\alpha}\tau^\dagger_{3\beta}\tau^\dagger_{4\beta} +
\tau_{1\alpha}\tau_{2\alpha}\tau_{3\beta}\tau_{4\beta}) +
\delta_{1+2-3-4} (\vqb
\tau^\dagger_{1\alpha}\tau^\dagger_{2\alpha}\tau_{3\beta}\tau_{4\beta}
+\vqc\tau^\dagger_{1\alpha}\tau^\dagger_{2\beta}\tau_{3\alpha}\tau_{4\beta})
\nonumber \\
&~~~~~~~~~~~+\delta_{1+2+3-4} \vqd
(\tau^\dagger_{1\alpha}\tau^\dagger_{2\alpha}\tau^\dagger_{3\beta}\tau_{4\beta} +
\tau^\dagger_{4\beta} \tau_{3\beta}\tau_{2\alpha}\tau_{1\alpha})\big]
\end{align}
where the momenta have been abbreviated according to ${\vec k}_1 \equiv 1$ etc.,
and the vertex functions $\vqa \ldots \vqd$ are given in Appendix~\ref{app:vertices}.
For $d=2$ our expressions (\ref{hp0}--\ref{hp4}) agree with those given in
Ref.~\onlinecite{alter2}.
Finally, the cubic term, present only in the asymmetric case $\lm\neq 0$, reads:
\begin{align}
\label{hp3}
\mathcal{H}'_3 = \frac{1}{\sqrt{N}} \sum_{123} \epsilon_{\alpha\beta\gamma}\big[
&\delta_{1+2+3}\vca (\tau^\dagger_{1\alpha}\tau^\dagger_{2\beta}\tau^\dagger_{3\gamma}- \tau_{1\alpha}\tau_{2\beta}\tau_{3\gamma})
+\delta_{1+2-3}\vcb (\tau^\dagger_{1\alpha}\tau^\dagger_{2\beta}\tau_{3\gamma}- \tau^\dagger_{3\gamma}\tau_{2\beta}\tau_{1\alpha})  \nonumber \\
+&\delta_{2+3-1}\vcc (\tau^\dagger_{3\gamma}\tau^\dagger_{2\beta}\tau_{1\alpha}- \tau^\dagger_{1\alpha}\tau_{2\beta}\tau_{3\gamma})
+\delta_{1-2+3}\vcd (\tau^\dagger_{1\alpha}\tau^\dagger_{3\gamma}\tau_{2\beta}- \tau^\dagger_{2\beta}\tau_{3\gamma}\tau_{1\alpha})
\big]\,,
\end{align}
with its vertex functions $\vca \ldots \vcd$ listed in Appendix~\ref{app:vertices}.

\end{widetext}


\section{$1/d$ expansion for observables}
\label{sec:expa}

As announced, we now evaluate important observables, organizing the perturbative contributions in an
expansion in $1/d$. Based on the Hamiltonian in Eq.~\eqref{hpexp}, diagrammatics is done using
$\mathcal{H}'_{2a}$ as unperturbed piece and $\mathcal{H}'_{2b}+\mathcal{H}'_{3}+\mathcal{H}'_{4}+\mathcal{H}'_{5}+\mathcal{H}'_{6}$ as
perturbation. The calculation will be limited to the leading corrections beyond the harmonic approximation -- as will become clear below, these corrections will enter at different orders in $1/d$ for different observables.

We exclusively consider $T=0$; this greatly reduces the number of contributing diagrams as all
closed (unidirectional) loops of $\tau$ particles vanish in the vacuum state.
Evaluating individual diagrams involving cubic or quartic vertices typically leads to a large number of terms, most of which turn out to {\em not} contribute to the leading $1/d$ corrections. In this section, we will restrict the presentation to quoting the relevant results; a more detailed exposure of how to extract a $1/d$ expansion can be found in Appendix~\ref{app:dgrdemo} for one sample diagram.

\subsection{Ground-state energy}

We start with the ground-state energy per dimer.
The harmonic-approximation result follows from $\mathcal{H}_0$ and $\mathcal{H}_2$
\eqref{hh2kd}:
\begin{equation}
\frac{E^{\rm harm}_0}{JN} = -\frac{3}{4} +
\frac{3}{2JN} \sum_{\vec k} (\wk - A_{\vec k}) ~\overset{d\to\infty}{=}~ -\frac{3}{4}-\frac{3}{8}\frac{\qq^2}{d}
\end{equation}
where the last expression involves an expansion to leading order in $1/d$ as described in
Appendix~\ref{app:expval}; it is identical to an expansion up to order $1/d$ of
$\mathcal{H}'_0$ in Eq.~\eqref{hp0}.

\begin{figure}[b!]
\includegraphics[width=0.47\textwidth]{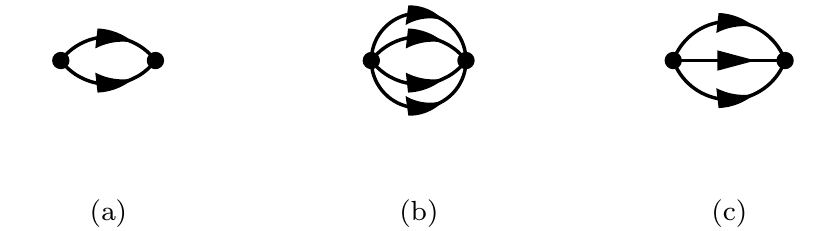}
\caption{ Feynman diagrams for the ground-state energy up to order $1/d^2$.}
\label{fig:dgre0}
\end{figure}

Higher-order terms involve the perturbative couplings and can be calculated diagrammatically. Up to order $1/d^2$ there are two diagrams contributing in the $\lm=0$ case, Figs.~\ref{fig:dgre0}(a,b), and one further diagram involving cubic vertices which are non-zero only for $\lm\neq0$, Fig.~\ref{fig:dgre0}(c).
The ground-state energy is then the sum of all these contributions,
$E_0 = \mathcal{H}'_0 + E_0^{\ref{fig:dgre0}(a)} + E_0^{\ref{fig:dgre0}(b)} + E_0^{\ref{fig:dgre0}(c)}$.
The diagram in Fig.~\ref{fig:dgre0}(a), being of second order in $\mathcal{H}'_{2b}$, evaluates to
\begin{equation}
E_0^{\ref{fig:dgre0}(a)}
= - 3 \sum_{\vec k} \frac{D_{\vec k}^2}{4\wk}\,.
\end{equation}
Given that the $D_{\vec k}$ vertex \eqref{eq:dk} is of order $1/d$, only those terms in
$E_0^{\ref{fig:dgre0}(a)}$ will contribute to order $1/d^2$ which are not further suppressed by the
momentum summation. This implies to approximate $\wk$ \eqref{om0} by its zeroth-order term in $\gk$, $\wk \approx J$, and leads to
\begin{equation}
\frac{E_0^{\ref{fig:dgre0}(a)}}{JN}
= - 3 \qq^2 (R_3+R_4)^2 = -\frac{3}{16} \frac{\qq^4}{d^2}
\end{equation}
to order $1/d^2$.

Turning to the second-order $\mathcal{H}'_{4}$ diagram, shown in Fig.~\ref{fig:dgre0}(b), we notice
that this has vertices $\vqa$ of order $1/d^0$, but will be suppressed at least down to $1/d^2$ by
internal momentum summations involving $\gk$ or $v_{\vec k}$ factors from the vertices.
Hence, the energies of the internal particle lines can again be approximated by $\wk \approx J$.
Enumerating all possible contractions of internal lines and using the explicit form of $\vqa$ we find
to order $1/d^2$:
\begin{equation}
\frac{E_0^{\ref{fig:dgre0}(b)}}{JN}
= -\frac{3}{8} \frac{\qq^4}{d^2}.
\end{equation}
Similarly, we find the contribution from the cubic diagram to order $1/d^2$:
\begin{equation}
\frac{E_0^{\ref{fig:dgre0}(c)}}{JN}
= - \frac{\lm^2 \qq^2}{3 J^2}\left(\frac{R_2}{2d}-R_{3}^{2}\right) = 0\,.
\end{equation}
This is an accidental cancellation, leading to a $\lm$-independent ground-state energy to order $1/d^2$. We do not expect such cancellations at higher orders, see also Eq.~\eqref{e0_smallx2} below.

Finally, we need the expansion of $\mathcal{H}'_0$ \eqref{hp0} to order $1/d^2$:
\begin{equation}
\frac{\mathcal{H}'_0}{JN}
= -\frac{3}{4} - \frac{3\qq^2}{8d} - \frac{3\qq^3}{16d^2} + \frac{27 \qq^4}{64d^2}.
\end{equation}
Collecting all terms gives our result for $E_0$:
\begin{equation}
\label{e0_larged}
\frac{E_0}{JN} = -\frac{3}{4}
-\frac{3}{8} \frac{\qq^2}{d} - \frac{3}{16} \frac{\qq^3}{d^2}  - \frac{9}{64}
\frac{\qq^4}{d^2} + \mathcal{O}\left(\frac{1}{d^3}\right) \,.
\end{equation}
The ground-state energy being analytic up to the critical point is consistent with the mean-field value\cite{goldenfeld} $\alpha=0$ for the specific-heat critical exponent $\alpha$.

\subsection{Triplet density}
\label{sec:expalocal}

We continue by calculating additional local static observables.
The local triplet density $\langle t_{i\alpha}^\dagger t_{i\alpha} \rangle$ per site vanishes as $d\to\infty$ as stated above; it also vanishes as $\qq\to 0$ for any $d$. In the harmonic approximation we have $\sum_i \langle t_{i\alpha}^\dagger t_{i\alpha} \rangle/N = R_2$, see Appendix~\ref{app:expval}.  Perturbative corrections, which can be calculated based on the $\tau$-particle self-energies described in more detail in the next subsection, start only at order $1/d^2$, such that we have:
\begin{equation}
\frac{1}{N} \sum_i \langle t_{i\alpha}^\dagger t_{i\alpha} \rangle = \frac{\qq^2}{8 d} + \mathcal{O}\left(\frac{1}{d^2}\right) \,.
\end{equation}
Notably, obtaining the complete $1/d^2$ contribution would require self-energies at next-to-leading order (i.e.~$1/d^2$) which are beyond the scope of this paper.

The expectation value of the bond-pair creation operator, $\sum_{\langle ij\rangle} \langle t_{i\alpha}^\dagger t_{j\alpha}^\dagger \rangle$, involves two different sites and hence an additional factor of $\gk$. As a result, we can obtain the full $1/d^2$ correction, with the following result for $\lm=0$:
\begin{equation}
\frac{1}{Nd} \sum_{\langle ij\rangle} \langle t_{i\alpha}^\dagger t_{j\alpha}^\dagger \rangle = -\frac{\qq}{4 d} - \frac{(2\qq^2+\qq^3)}{16d^2} + \mathcal{O}\left(\frac{1}{d^3}\right)\,.
\end{equation}

Finally, it is instructive to consider the site-pair creation operator, $\langle t_{i\alpha}^\dagger t_{i\alpha}^\dagger \rangle$ -- this quantity must vanish as a result of the constraint \eqref{hardcore}. In the harmonic approximation we have $\sum_i \langle t_{i\alpha}^\dagger t_{i\alpha}^\dagger \rangle/N = R_1$, but perturbative corrections start at order $1/d$ and cancel the harmonic result, such that eventually $\langle t_{i\alpha}^\dagger t_{i\alpha}^\dagger \rangle = 0$ to order $1/d$; we expect this to hold order by order in the $1/d$ expansion.\cite{anomfoot}

\subsection{Triplon dynamics}
\label{sec:expadisp}

The leading-order triplon dispersion $\wk$ is from the harmonic approximation, with the result in
Eq.~\eqref{om0}. Corrections from the perturbative couplings can be evaluated via self-energies which,
in the large-$d$ limit, are all suppressed at least as $1/d$. Importantly, we will have both normal and anomalous $\tau$ diagrams, such that the Dyson equation takes the following form:
\begin{align}
\label{eq:green1}
\mathcal{G}^{N}(\vec{k},\w)&= \frac{\w + \wk + \Sigma_{N}(\vec{k},-\w)} {\Xi(\w,\vec{k})}\,, \\
\mathcal{G}^{A}(\vec{k},\w)&= \frac{-\Sigma_{A}(\vec{k},\w)} {\Xi(\w,\vec{k})}
\end{align}
with
\begin{align}
\label{eq:pole1}
\Xi(\w,\vec{k}) &= \big[\w+\wk+\Sigma_{N}(\vec{k},-\w)\big] \big[ \w-\wk-\Sigma_{N}(\vec{k},\w)\big] \notag \\ &~~~~~~~~~~~~~~~~~~~~~~+ \Sigma_{A}(\vec{k},\w) \Sigma_{A}(\vec{k},-\w)\,.
\end{align}
Consequently, the equation for the renormalized pole energies $\Wk$ is $\Xi(\Wk,\vec{k})=0$.

In general, the self-energies $\Sigma_{N,A}$ entering $\Xi(\vec k,\w)$ need to be evaluated at $\w=\Wk$; for the $1/d$ expansion this means that the energy argument of $\Sigma_{N,A}$ itself needs to be expanded in $1/d$, according to:
\begin{equation}
\label{sigma-exp}
\Sigma_{N}(\vec{k},\pm \Wk) = \Sigma_{N\pm} + (\Wk - \wk) \Sigma'_{N\pm}\,,
\end{equation}
with the abbreviations
\begin{align}
\label{sigma-p1}
\Sigma_{N\pm} = \Sigma_{N}(\vec{k},\pm \wk),~~
\left. \Sigma'_{N\pm} = \frac{\partial \Sigma_{N}(\vec{k},\pm \w)}{\partial \w} \right|_{\w=\wk}
\!\!\!\!\!.
\end{align}
In the following, we calculate $\Wk$ up to order $1/d$ only. This leads to two simplifications: The
self-energies can be evaluated at the unperturbed $\wk$, $\Sigma_{N}(\vec{k},\pm \Wk) \approx \Sigma_{N\pm}$, and the $\Sigma_{N,A}^2$ terms in Eq.~\eqref{eq:pole1} can be neglected as they are of order $1/d^2$.
These simplifications reduce the (positive-energy) pole equation to
\begin{equation}
\Wk - \wk - \Sigma_{N+} = 0\,.
\end{equation}
As discussed in Section~\ref{sec:mfchar}, expansions have to be used with care in the vicinity of the quantum critical point. In particular, $\Wk$ will not have a well-defined $1/d$ expansion near $\vec k=\vec Q$ when the gap closes. However, $\Wk^2$ can be expected to be analytic for the same reason as $\Delta^2 \propto (q_c - q)^{2\nu z} = q_c-q$ is analytic.
Consequently, we shall work with the following dispersion expression, valid to order $1/d$:
\begin{equation}
\label{pole}
\Wk^{2} = \wk^{2} + 2 \wk \Sigma_{N+}\,.
\end{equation}

\begin{figure}
\includegraphics[width=0.47\textwidth]{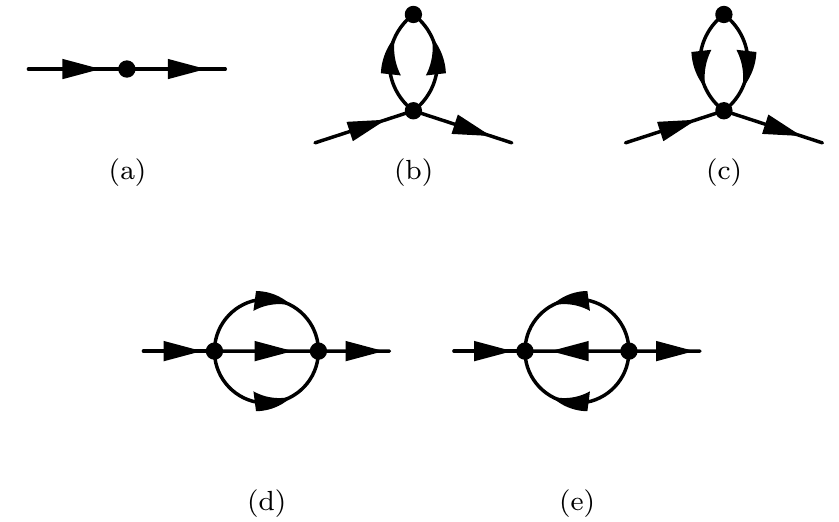}
\caption{Feynman diagrams for the normal $\tau$ self-energy up to order $1/d$, with vertices from $\mathcal{H}'_{2b}$ and $\mathcal{H}'_4$.}
\label{fig:dis}
\end{figure}

The diagrams contributing to the normal self-energy in the symmetric case, $\lm=0$, and to order $1/d$ are shown in Fig.~\ref{fig:dis}.
Evaluating the frequency and momentum integrals, again approximating the energies of the internal
particle lines by $\wk \approx J$, we find the following contributions to order $1/d$ (see Appendix~\ref{app:dgrdemo} for a guide):
\begin{align}
\label{bilinear}
\Sigma^{3(a)}(\vec{k},\w) &= C_{\vec{k}}\,,
\end{align}
\begin{widetext}
\begin{align}
\label{quartic1}
\Sigma^{3(b)}(\vec{k},\w) &= \Sigma^{3(c)}(\vec{k},\w) = -\gk \qq^{2} J (R_3 + R_4) (u_{\vec{k}} + v_{\vec{k}})^{2} \,,
 \\
\label{quartic3}
\Sigma^{3(d)}(\vec{k},\w) &= \frac{\qq^{2} J^{2}}{\w-3J} \left[ 4 \gk^{2} (u_{\vec{k}} + v_{\vec{k}})^{2} R_2
+ 8 \gk (u_{\vec{k}}^{2} + u_{\vec{k}} v_{\vec{k}}) R_3 + \frac{2 u_{\vec{k}}^{2}}{d} \right] \,, \\	
\label{quartic4}
\Sigma^{3(e)}(\vec{k},\w) &= \frac{-\qq^{2} J^{2}}{\w+3J} \left[ 4 \gk^{2} (u_{\vec{k}} + v_{\vec{k}})^{2} R_2
+ 8 \gk (v_{\vec{k}}^{2} + u_{\vec{k}} v_{\vec{k}}) R_3 + \frac{2 v_{\vec{k}}^{2}}{d} \right] \,. 
\end{align}
\end{widetext}

The $\Sigma$ expressions above can be evaluated using the explicit large-$d$ expressions for the
Bogoliubov coefficients and the $R_{1\ldots4}$ in Appendix~\ref{app:expval}. Collecting all
contributions, we finally find the $1/d$ expansion of the triplon dispersion for $\lm=0$:
\begin{equation}
\label{Wkexp}
\frac{\Wk^{2}}{J^2}  = 1 + 2\gamma_{\vec k} \qq + \frac{1}{d} (2 \qq^{2} - \gk^{2} \qq^{3}) +
\mathcal{O}\left(\frac{1}{d^2}\right) \,.
\end{equation}
We see that interactions generically increase the triplon energy (for $\qq<2$ which holds everywhere in the disordered phase treated here) -- this is of course expected for dominantly repulsive quartic interactions.
While Eq.~\eqref{Wkexp} could in principle be converted into an expansion for $\Wk/J$, such a conversion is well-defined only if $\wk\neq0$, i.e., it fails for $\vec{k}=\vec{Q}$ at criticality, as anticipated.

\begin{figure}[!b]
\includegraphics[width=0.47\textwidth]{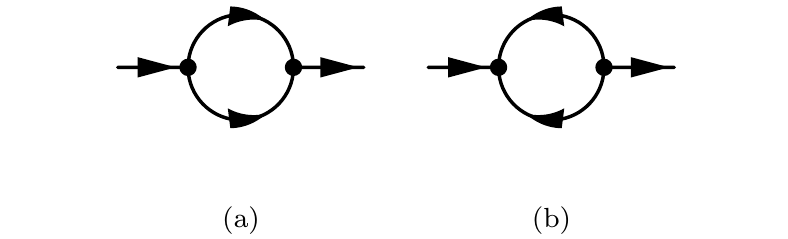}
\caption{Feynman diagrams for the contribution of cubic ($\mathcal{H}'_3$) terms to the normal $\tau$ self-energy up to order $1/d$.}
\label{fig:disc}
\end{figure}
In the asymmetric case, $\lm\neq 0$, additional self-energy diagrams involving cubic vertices occur; those are shown in Fig.~\ref{fig:disc}.
These diagrams are evaluated using the same prescription as discussed above for the quartic terms and explained in some detail in Appendix~\ref{app:dgrdemo}. To order $1/d$, these diagrams have following contributions:
\begin{widetext}
\begin{align}
\label{cubic1}
\Sigma^{4(a)}(\vec{k},\w) &= \frac{2 \lm^2 \qq^2 J^2}{\w-2J}\left\{ \frac{u_{\vec{k}}^2 (1-\gk)}{2d} + 2\gk(u_{\vec{k}}^2 + u_{\vec{k}}v_{\vec{k}})\left[R'_3(\vec{k}) - R_3\right]
+ \gk^2(u_{\vec{k}} + v_{\vec{k}})^2\left[R_2 - R'_{5}(\vec{k})\right]\right\} \\
\label{cubic2}
\Sigma^{4(b)}(\vec{k},\w) &= \frac{-2 \lm^2 \qq^2 J^2}{\w+2J}\left\{ \frac{v_{\vec{k}}^2 (1-\gk)}{2d} + 2\gk(v_{\vec{k}}^2 + u_{\vec{k}}v_{\vec{k}})\left[R'_3(\vec{k}) - R_3\right]
+ \gk^2(u_{\vec{k}} + v_{\vec{k}})^2 \left[R_2 - R'_{5}(\vec{k})\right]\right\},
\end{align}
see Appendix~\ref{app:expval} for $R'_5$. These self-energy contributions modify the triplon dispersion as follows:
\begin{equation}
\label{Wkexp-a}
\frac{\Wk^{2}}{J^2}  = 1 + 2\gamma_{\vec k} \qq + \frac{1}{d} (2 \qq^{2} - \gk^{2} \qq^{3}) + \frac{\lm^2 \qq^2 (1-\gk)(6+14\gk\qq+6\gk^2\qq^2)}{(2\gk\qq-3)d}
+\mathcal{O}\left(\frac{1}{d^2}\right) \,.
\end{equation}
\end{widetext}
This explicitly shows that the $1/d$ expansion is not simply an expansion in powers of $\qq$ or $\gk$: The non-trival denominator in the $\lm^2/d$ correction of Eq.~\eqref{Wkexp-a} arises as a product of the denominators in the self-energies \eqref{cubic1}
and \eqref{cubic2}, evaluated at $\w=\wk$. (In the symmetric case $\lm=0$ such denominators were cancelled by identical factors in the numerator.)

\begin{figure}[!h]
\includegraphics[width=0.48\textwidth]{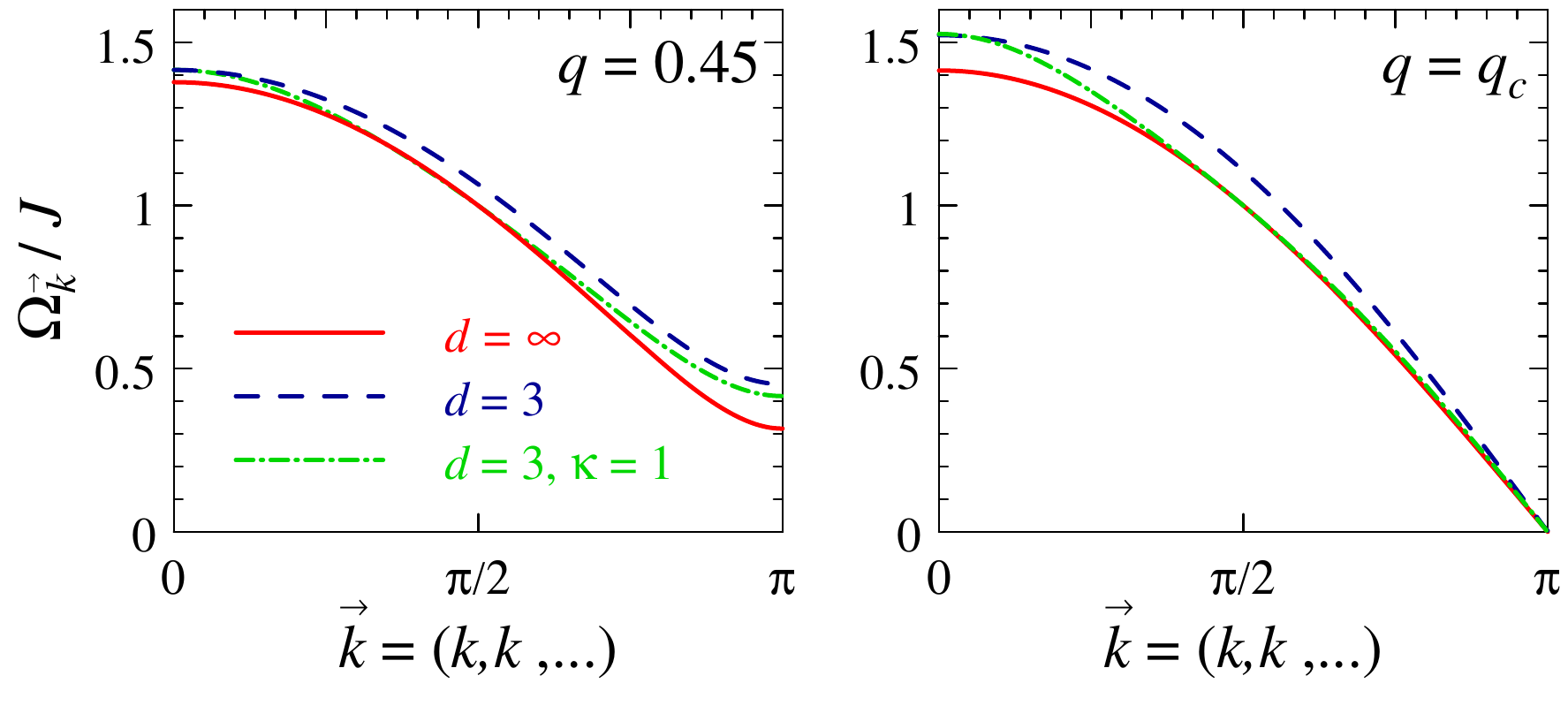}
\caption{
Triplon dispersion \eqref{Wkexp-a} derived from the $1/d$ expansion for the coupled-dimer model \eqref{hh}, showing results for $d=\infty$ (solid), $d=3$ with $\lm=0$ (dashed), and $d=3$ with $\lm=1$ (dash-dot), where $\lm$ is the asymmetry parameter \eqref{lamdef}.
Left: results for fixed $\qq=Kd/J=0.45$.
Right: results at criticality, $\qq=\qqc$, where here $\qqc$ is defined by $\W_{\vec Q}=0$ with $\Wk$ from Eq.~\eqref{Wkexp-a} at {\em fixed} $d$; the value of this $\qqc$ is distinct from the expansion result \eqref{qqc2} evaluated at fixed $d$.
}
\label{fig:disp1}
\end{figure}

We note that, to all orders, the momentum dependence enters via the structure factor $\gamma$ \eqref{gammadef} only, but at higher orders $\gamma_{2\vec k}$, $\gamma_{3\vec k}$ etc. may appear as well.
The dispersion results, for concrete values of $d$, are illustrated in Fig.~\ref{fig:disp1}.

\subsection{Triplon decay}

As can be seen from the explicit expressions, the self-energies are purely real for the relevant frequencies, i.e., there is no triplon damping. In fact, this result is not restricted to order $1/d$: In the large-$d$ limit, the typical $\gk$ is small, such that the triplon density of states is strongly peaked at $\w=J$. More precisely, the density of states for $\w\neq J$ is exponentially small\cite{dmft} as $d\to\infty$.
Consequently, the same applies to the density of states of multi-triplon continua which are responsible for damping, such that all damping rates (inverse lifetimes) are exponentially small in $1/d$ and thus vanish to all orders in a $1/d$ expansion.

We note that the poles in the self-energies, located at $\pm 2J$ and $\pm 3J$ at order $1/d$, produce additional spectral weight in the triplon propagators near these frequencies. This weight takes the form of poles with strengths of order $1/d$, which mimic the incoherent continuum present at finite $d$.

\subsection{Gap and phase boundary}

\begin{figure}[!t]
\includegraphics[width=0.43\textwidth]{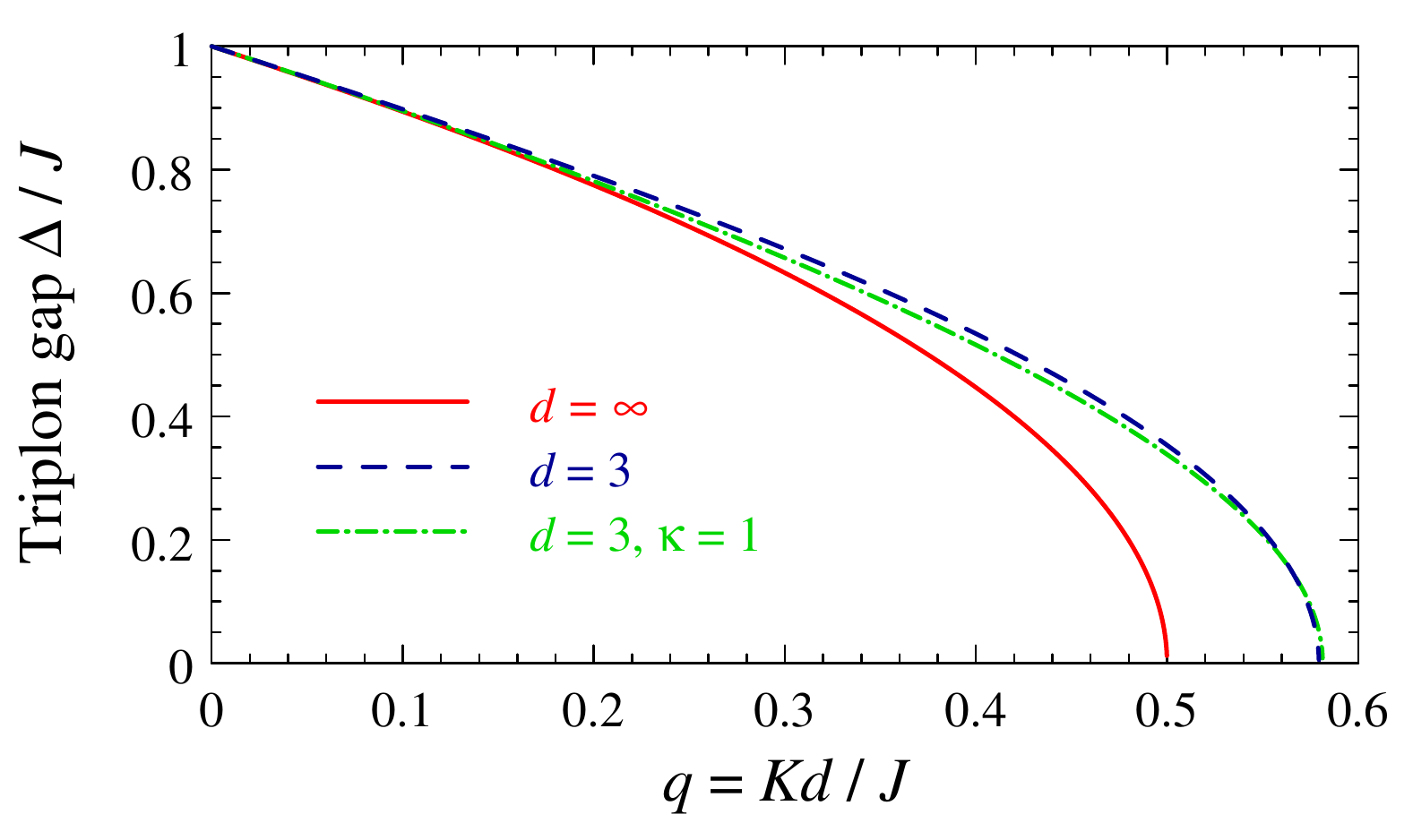}
\caption{
Triplon gap \eqref{gap-a} for $d=\infty$ (solid), $d=3$ with $\lm=0$ (dashed), and $d=3$ with $\lm=1$ (dash-dot).
}
\label{fig:gap1}
\end{figure}

The excitation gap of the paramagnetic phase, $\Delta$, is simply given by the minimum of the triplon dispersion, $\Delta = \W_{\vec Q}$. This yields
\begin{widetext}
\begin{equation}
\frac{\Delta^{2}}{J^2} = 1-2\qq + \frac{1}{d} (2 \qq^{2} -  \qq^{3})  - \frac{2\lm^2 \qq^2(6-14\qq+6\qq^2)}{(2\qq+3)d}
+\mathcal{O}\left(\frac{1}{d^2}\right)
\label{gap-a}
\end{equation}
and is graphically shown in Fig.~\ref{fig:gap1}.
As announced in Section~\ref{sec:mfchar}, we find an expansion for $\Delta^2$ which is well-behaved even at criticality; this would not apply to $\Delta$.
Near $\vec Q$ we can expand $\gk \approx -1 + \sum_n (k_n - \pi)^2 / (2d)$. This
yields the parametrization in Eq.~\eqref{quadpara}, with the mode velocity $c$ given by
\begin{equation}
\label{callq}
\frac{c}{J} = \sqrt{\qq} + \frac{\qq^{5/2}}{2 d} -\frac{\lm^2\qq^{3/2}}{2(2\qq+3)d}\left[
\frac{(6-14\qq+6\qq^2)(2\qq-3)}{2(2\qq+3)}+14\qq-12\qq^2\right]+ \mathcal{O}\Big(\frac{1}{d^2}\Big) .
\end{equation}
\clearpage
\end{widetext}

The location $\qqc$ of the boundary to the magnetically ordered phase can be obtained from the
condition \mbox{$\Delta^2(\qqc) = 0$}. Using an ansatz $\qq_c = 1/2 + \qq_{c1}/d$ we can obtain $\qq_{1c}$ and with it the phase boundary to order $1/d$:
\begin{equation}
 \qq_c = \frac{1}{2} + \left(\frac{3}{16}+\frac{\lm^2}{32}\right)\frac{1}{d} + \mathcal{O}\left(\frac{1}{d^2}\right)\,.
\label{qqc2}
\end{equation}
For $\lm=0$, this reduces to the result announced in the introduction, Eq.~\eqref{qqcexp}.
The gap $\Delta$ vanishes in a square-root fashion upon approaching $\qqc$. Extracting the prefactor of the square root yields the result \eqref{deltanearqc}.

Evaluating the expression \eqref{qqc2} for $d=2$ and $\lm=0$ yields a result for the critical coupling of the Heisenberg bilayer model significantly smaller than the value known from QMC calculations,\cite{sandvik06} see Fig.~\ref{fig:pd1}. This indicates sizeable contributions from higher orders in the $1/d$ expansion; we leave their explicit evaluation for future work.

It is worth noting that solving the equation $\Delta^2(\qqc) = 0$ using the truncated series \eqref{gap-a} for finite $d$ yields a value for $\qqc$ which is distinct from $\qqc$ as given by the truncated series \eqref{qqc2} for the same finite $d$. The reason is simply that $\Delta^2$ from Eq.~\eqref{gap-a} evaluated at $\qqc$ from Eq.~\eqref{qqc2} contains $1/d^2$ terms which do not vanish, see also Fig.~\ref{fig:pd1}.


\subsection{Triplon weight in dynamic susceptibility}

To complete the analysis, we determine the weight of the triplon mode in the dynamic spin susceptibility,
\begin{equation}
\label{susc}
\chi_{\alpha}(\vec{k},\w) = -\ii \int_{-\infty}^{\infty} dt e^{\ii \w t} \langle T_t S_{\alpha}(\vec{k},t) S_{\alpha}(-\vec{k},0) \rangle,
\end{equation}
restricting the analysis to the symmetric case, $\lm = 0$.
In the coupled-dimer system under consideration, the fourier-transformed spin operator $S_{\alpha}(\vec{k})$ has two contributions with different form factors, namely even ($e$) and odd ($o$) ones:
\begin{align}
S_{\alpha}^{e}&= S_{\alpha}^1 + S_{\alpha}^2 = -\ii \epsilon_{\alpha\beta\gamma} t_{\beta}^{\dagger}t_{\gamma} \, \\
S_{\alpha}^{o}&= S_{\alpha}^1 - S_{\alpha}^2 = t_{\alpha}^{\dagger} P + P t_{\alpha} \,,
\end{align}
with $P$ the projector of Eq.~\eqref{linproj}.
To extract the mode weight we restrict our attention to those contributions to $\chi_{\alpha}(\vec{k},\w)$ which correspond to a single-mode response, i.e., we do not consider the even channel which produces a two-particle continuum only. We note, however, that the $P$ in $S_{\alpha}^{o}$ influences the pole weight at order $1/d$ in a non-trivial fashion and can be approximated neither by unity nor by $\langle P \rangle$.

Using the Bogoliubov transformation \eqref{bogol} one can write the spin susceptibility for $S_{\alpha}^{o}$ to order $1/d$ in terms of the
$\tau$-Green's functions as follows:
\begin{align}
\tilde{\chi}_{\alpha}(\vec{k},\w) &= (u_{\vec{k}} + v_{\vec{k}})^2 (1 - 2 R_{1} - 8 R_{2})
\Big[ \mathcal{G}^{N}(\vec{k},\w) \nonumber \\
&+ \mathcal{G}^{N}(\vec{k},-\w) + \mathcal{G}^{A}(\vec{k},\w) + \mathcal{G}^{A}(\vec{k},-\w)
\Big] \,.
\end{align}
Since we are interested in the pole weight, we need to analyze $\tilde{\chi}$ in the vicinity of the pole at $\Wk$. Expanding the self-energies in the vicinity of $\w=\wk$ and using the relations
\eqref{sigma-exp} and \eqref{pole} we can cast the Green's functions into the following form:
\begin{align}
\mathcal{G}^{N}(\vec{k},\w) &= \frac{(1-\Sigma'_{N+})^{-1}}{\w-\Wk},   \\
\mathcal{G}^{N}(\vec{k},-\w) &= - \frac{(1-\Sigma'_{N-})^{-1}}{\w+\Wk^{-}}, \\ 
\mathcal{G}^{A}(\vec{k},\w) &= -\frac{\Sigma_{A+} + (\w-\wk)\Sigma'_{A+}}{(\w-\Wk)(\w+\Wk^{-})}, \\ 
\mathcal{G}^{A}(\vec{k},-\w) &= -\frac{\Sigma_{A-} + (\w-\wk)\Sigma'_{A-}}{(\w-\Wk)(\w+\Wk^{-})},  
\end{align}
valid to order $1/d$. Here we have used the abbreviations $\Sigma_{N\pm}$ and $\Sigma'_{N\pm}$ of Eq.~\eqref{sigma-p1} and similar ones for the anomalous self-energy, and we have
defined
\begin{equation}
\Wk^{-} = \wk(1-2\Sigma'_{N-})+\Sigma_{N-}.
\end{equation}
Additionally, for the anomalous self-energy, we have $\Sigma_{A+} = \Sigma_{A-}$ and $\Sigma'_{A+} = \Sigma'_{A-}$ to order $1/d$.
The susceptibility in the vicinity of $\w=\wk$ becomes
\begin{widetext}
\begin{align}
\tilde{\chi}_{\alpha}(\vec{k},\w) &= (u_{\vec{k}} + v_{\vec{k}})^2  (1 - 2R_{1} - 8R_{2}) \bigg\lbrace
\frac{1}{\w-\Wk}\left[(1-\Sigma'_{N+})^{-1} - 2\frac{\Sigma_{A+}- \wk \Sigma'_{A+}+ \Wk \Sigma'_{A+}}{\Wk+\Wk^{-}}\right] \nonumber \\
&-\frac{1}{\w+\Wk^{-}}\left[(1+\Sigma'_{N-})^{-1} - 2\frac{\Sigma_{A+}- \wk \Sigma'_{A+}- \Wk^{-} \Sigma'_{A+}}{\Wk+\Wk^{-}}\right]
\bigg\rbrace
\end{align}
\end{widetext}
It is then easy to identify the pole weight corresponding to $\Wk$ as
\begin{equation}
\mathcal{Z}_{\vec{k}} = (u_{\vec{k}} + v_{\vec{k}})^2
\left[1 + \Sigma'_{N+}-\frac{\Sigma_{A+}}{\wk} - 2R_{1} - 8R_{2} \right]
\end{equation}
where $\Sigma_{-}$ and $\Sigma'_{-}$ have disappeared, as they characterize the self-energy away from the pole.

\begin{figure}[b]
\includegraphics[width=0.47\textwidth]{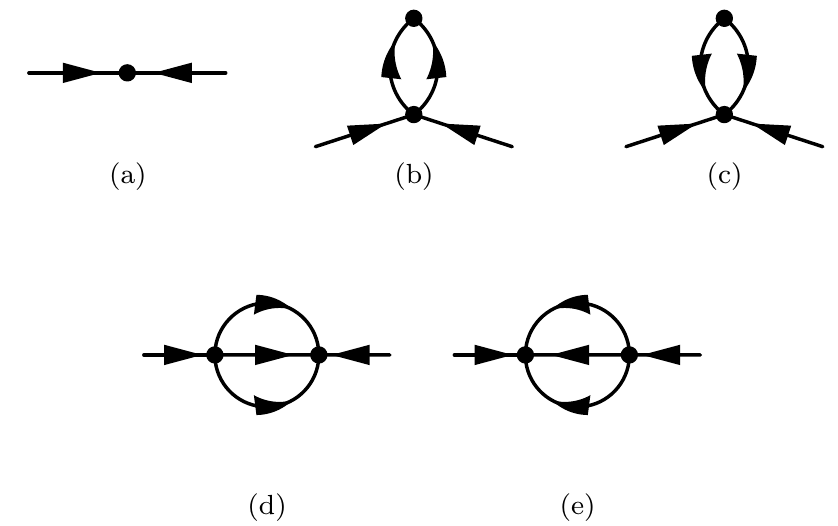}
\caption{Feynman diagrams for anomalous self-energies contributing to order $1/d$.}
\label{fig:anomal}
\end{figure}

To obtain an explicit expression for the pole weight we also need the contributions to the anomalous self-energy, with the relevant diagrams shown in Fig.~ \ref{fig:anomal}.
Their evaluation can be done along the lines discussed in the previous subsections, with the following results:
\begin{align}
\label{anomal1}
\Sigma^{\ref{fig:anomal}(a)} &= D_{\vec{k}}, \\
\Sigma^{\ref{fig:anomal}(b)} &= -2\gk \qq^{2} J R_{3} (u_{\vec{k}}^{2} + u_{\vec{k}} v_{\vec{k}}), \\ 
\Sigma^{\ref{fig:anomal}(c)} &= -2\gk \qq^{2} J R_{3} (v_{\vec{k}}^{2} + u_{\vec{k}} v_{\vec{k}}), \\ 
\Sigma^{\ref{fig:anomal}(d)} &= \frac{\qq^{2} J^2}{\w-3J} \left[ 4 \gk (u_{\vec{k}} + v_{\vec{k}})^{2} (\gk R_2 + R_3) + \frac{2 u_{\vec{k}}v_{\vec{k}}}{d} \right],\\
\Sigma^{\ref{fig:anomal}(e)} &= \frac{-\qq^{2} J^2}{\w+3J} \left[ 4 \gk (u_{\vec{k}} + v_{\vec{k}})^{2} (\gk R_2 + R_3) + \frac{2 u_{\vec{k}}v_{\vec{k}}}{d} \right],
\end{align}
where the self-energy arguments $(\vec k,\w)$ have been omitted.
Inserting the expressions of the self-energies evaluated here and in Eqs.~\eqref{bilinear}--\eqref{quartic4}, we obtain for the pole weight:
\begin{equation}
\mathcal{Z}_{\vec{k}} = \frac{J}{\wk}\left\lbrace 1 - \frac{\qq^2}{2d}\left[ 7
+ \frac{1+\gk-2\gk\qq+\gk^2\qq}{1+2\gk\qq}\right] \right\rbrace.
\end{equation}
This expression is seen to be singular at the bare critical point, i.e., $\qq=1/2$ and $\gk=-1$.
However, this singularity can be removed by realizing that the physical pole weight will diverge for $\Wk\to 0$ (instead of $\wk\to 0$). Hence, upon casting the above expression into the form
\begin{equation}
\mathcal{Z}_{\vec{k}} = \frac{J}{\Wk}\mathcal{W}_{\vec{k}}\,,
\end{equation}
the prefactor $\mathcal{W}_{\vec{k}}$ is expected to have a regular $1/d$ expansion.
Indeed, after a few steps of algebra one finds:
\begin{equation}
\mathcal{W}_{\vec{k}} = 1- \frac{\qq^2}{2d} (6+\gk) +\mathcal{O}\left(\frac{1}{d^2}\right)\,,
\end{equation}
which constitutes our final result for the $\lm=0$ susceptibility pole weight.


\section{Inter-dimer perturbation theory for arbitrary $d$}
\label{sec:smallk}

In this section, we turn to employing an entirely different method to calculate observables for the
hypercubic coupled-dimer model \eqref{hh}, namely a high-order series expansion in the relative
strength of the inter-dimer coupling, $\x = K/J$. Expansions of this type have been used
before for lattices in $d=1$ and $2$;\cite{Weihong97,Knetter03_2} here we will derive
results for arbitrary spatial dimension $d$.
We note that high-temperature expansions for Ising and Potts models
on the hypercubic lattice have been performed in Refs.~\onlinecite{Fisher90, Meik06}, but we are not aware of any such expansions for one-particle dispersions in a quantum lattice problem. Below, we shall use the results as an independent check of our $1/d$ expansion developed in this paper. Furthermore, such high-order series expansions for general $d$ represent an interesting tool to investigate quantum phase transitions; we will address this aspect in a forthcoming publication.\cite{Coester14}

\subsection{Method}

We start by sketching the methodology of the expansion; for details we refer the reader to Refs.~\onlinecite{gsu00,Knetter03_1}.
The expansion's reference point corresponds to $\x=0$. Here the ground state is given by a product state of singlets on the dimers, and elementary excitations are local triplets with excitation energy $\Delta=J$. After a global energy shift, we can rewrite Hamiltonian (\ref{hh}) in the form
\begin{eqnarray}
\label{h_pert}
\mathcal{H}&=&\mathcal{H}_0+k\, \hat{V} \quad ,
\end{eqnarray}
where $\mathcal{H}_0$ has an equidistant spectrum bounded from below counting the number of triplets. Furthermore, the perturbing part can be written as
\begin{align}
\hat{V}=\hat{T}_{-2}+\hat{T}_{-1}+\hat{T}_0+\hat{T}_{1}+\hat{T}_{2} \, ,
\end{align}
where $\hat{T}_m$ changes the total number of triplets by \mbox{$m\in\{\pm 2,\pm 1, 0\}$}. Note that terms with odd $m$ only appear in the asymmetric case, $\lm\neq 0$, corresponding to cubic terms in the bond-operator language.

Each operator $\hat{T}_m$ is a sum over local operators connecting two nearest-neighbor dimers. One can therefore write
\begin{align}
\hat{T}_m=\sum_l \hat{\tau}_{m,l}\,,\label{tausumme}
\end{align}
with $\hat{\tau}_{m,l}$ effecting only the two dimers connected by the link $l$ on the lattice.

The perturbative continuous unitary transformations (pCUTs) \cite{gsu00,Knetter03_1} map the original Hamiltonian to an effective quasiparticle conserving Hamiltonian of the form
\begin{eqnarray}
\hat{H}_\text{eff}(k)=\hat{H}_0+\sum_{n=1}^{\infty}\x^n \hspace*{-2mm}
\sum_{{\rm dim}(\underline{m})=n \atop  \,M(\underline{m})=0} \hspace*{-2mm}
C(\underline{m})\,\hat{T}_{m_1}\dots \hat{T}_{m_n},
\end{eqnarray}
where $n$ reflects the perturbative order. The second sum is taken over all possible vectors $\underline{m}\!\equiv\!(m_1, \ldots, m_n)$ with $m_i\in\{\pm 2,\pm 1,0\}$ and dimension ${\rm dim}(\underline{m})=n$. Each term of this sum is weighted by the rational coefficient \mbox{$C(\underline{m})\in \mathbb{Q}$} which has been calculated model-independently up to high orders.\cite{gsu00} The additional restriction $M(\underline{m})\equiv\sum m_i=0$ reflects the quasiparticle-conserving property of the effective Hamiltonian, i.e., the resulting Hamiltonian is block-diagonal in the number of quasiparticles $[\hat{H}_\text{eff},\hat{H}_0]=0$. Each quasiparticle block can then be investigated separately which represents a major simplification of the complicated many-body problem.

The operator products $\hat{T}_{m_1}\dots \hat{T}_{m_n}$ appearing in order $n$ can be interpreted as virtual fluctuations of ``length'' $l\leq n$ leading to dressed quasiparticles. According to the linked-cluster theorem, only linked fluctuations can have an overall contribution to the effective Hamiltonian $\hat{H}_\text{eff}$. Hence, the properties of interest can be calculated in the thermodynamic limit by applying the effective Hamiltonian on finite clusters.

Considering all linked fluctuations on the lattice (for arbitrary $d$), it becomes clear that the contribution of each fluctuation only depends on its topology.
We can therefore perform our calculations only on a finite set of topologically distinct graphs. The contribution on the graphs has then to be embedded into the lattice in order to extract the properties in the thermodynamic limit. In the following this is done for the ground-state energy and the one-triplon dispersion.

\subsection{Ground-state energy}

We now calculate the ground-state energy $E_0$ of the hypercubic-lattice coupled-dimer model for arbitrary $d$ up to order $\x^7$, using pCUTs and a full graph decomposition. This task is achieved in two steps: (i) extracting the ground-state energy per dimer on each graph in order seven and (ii) embedding these graph contributions into the lattice and summing up their contributions.

The first step is conventional and it is part of any linked-cluster expansion. In order to avoid double counting of contributions, the reduced contribution $\epsilon_{0,n}$ to $E_0$ of each graph $\mathcal{G}_n$ has to be calculated by subtracting the contributions of all subgraphs.

Up to order $n$, only graphs up to $n$ links have to be considered due to the linked-cluster theorem.
Now one has to check whether the graphs fit onto the lattice and whether each graph has a finite contribution in the order under consideration. The latter depends on both the model and the observable. In the case of the ground-state energy of the hypercubic-lattice coupled-dimer model, one has a specific selection rule that each link has to be touched twice by the perturbation as long as it is not part of a closed loop of links. This property drastically reduces the total number of graphs which one has to treat.
The relevant graphs for the calculation of the ground-state energy per dimer up to order seven are $\mathcal{G}_1$, $\mathcal{G}_2$, $\mathcal{G}_3$, $\mathcal{G}_4$, $\mathcal{G}_5$, $\mathcal{G}_7$, $\mathcal{G}_8$ and $\mathcal{G}_9$ which are all illustrated in Fig.\ref{fig:graphs_1qp}. Other graphs like $\mathcal{G}_6$, $\mathcal{G}_{10}$, and $\mathcal{G}_{11}$ do not contribute up to this order due to the double-touch property.

The embedding factor $\nu_n(d)$ for graph $\mathcal{G}_n$, being the number of possible embeddings of $\mathcal{G}_n$ on the lattice, is a function of the spatial dimension $d$. The ground-state energy per dimer in the thermodynamic limit is then given by
\begin{align}
\frac{E_0}{JN}=\sum_n \nu_n(d)\, \epsilon_{0,n}\,.
\end{align}
The determination of the embedding factors $\nu_n (d)$ for arbitrary $d$ is the most challenging part of this calculation.

In order to determine the embedding factors it is necessary to divide the number of naive embeddings by the symmetry factor $S_n$ of $\mathcal{G}_n$. Otherwise one overcounts contributions, since embeddings connected by a symmetry-mapping of the graph represent exactly the same fluctuation on the lattice in the thermodynamic limit.

Let us demonstrate the embedding procedure for graph $\mathcal{G}_3$. Without loss of generality, we can start the embedding from the dimer site $s_0$. Then, the site $s_1$ can be embedded in $2d$ possible directions; the site $s_2$ can be embedded in $(2d-1)$ possible directions because one direction is already occupied by $s_0$. The site $s_3$ can be embedded in $(2d-1)$ different directions because one direction is already occupied by $s_1$. Note that no possible direction is occupied by $s_0$ as the minimal loop in the hypercubic lattice is of length four. The symmetry factor is given by $S_3=2$ originating from a single reflection symmetry. We therefore end up with $\nu_3(d)=2d (2d-1)(2d-1)/2=d (2d-1)(2d-1)$.

Following these principles, we find the following small-$k$ expansion for the ground-state energy:
\begin{align}
\label{e0_smallx}
\frac{E_0}{JN} &= -\frac{3}{4}
-\frac{3}{8}d\,{\x}^{2}-\frac{3}{16}d\,{\x}^{3}
+\Big(\frac{21}{128}d-\frac{9}{64}d^2\Big){\x}^{4}\notag\\
&+\Big(\frac {57}{256}d-\frac{3}{64}d^2\Big){\x}^{5}
+\Big(-\frac {2781}{1024}d -\frac {7}{256}{\lm}^{2}d \notag\\
&~~~~~+ \frac {273}{64}d^2+\frac {7}{128}{\lm}^{2}d^2 -\frac{357}{256}d^3-\frac{1}{32}{\lm}^{2}d^3 \Big){\x}^{6} \notag\\
&+\Big(-\frac {73293}{16384}d -\frac {353}{1024}d{\lm}^{2} + \frac {53205}{8192}d^2+\frac {899}{1536}d^2{\lm}^{2} \notag\\
&~~~~~- \frac{8499}{4096}d^3 -\frac {97}{384}d^3{\lm}^{2} \Big){\x}^{7} +\mathcal{O}(\x^8)\,.
\end{align}
For $d=1$ and $\lm=0$ this formula reduces to the known results for the two-leg Heisenberg ladder.\cite{Knetter03_2} For the specific case $d=2$ and $\lm=0$, we reproduce the numerical results of the ground-state energy of the square-lattice bilayer.\cite{Weihong97}

\subsection{Triplon dispersion}

\begin{figure}[!tb]
\begin{center}
\includegraphics[width=6.8cm]{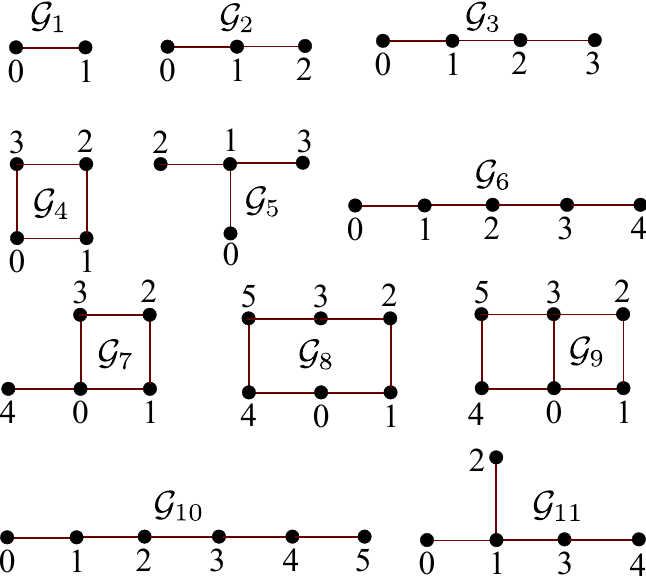}
\caption{Relevant graphs $\mathcal{G}_i$ with $i\in\{1,\ldots,11\}$ to calculate the one-particle dispersion up to order five in $\x$ and the ground-state energy per site up to order seven in $\x$ respectively. Here filled circles represent dimers which are connected their nearest-neighbors by links shown as solid lines.}
\label{fig:graphs_1qp}
\end{center}
\end{figure}

In this subsection we follow the same line as for the ground-state energy, but now calculate the one-particle dispersion $\Wk$ up to order $\x^5$. The dispersion is calculated by first determining the reduced one-particle hopping elements $t^{(n)}_{i,j}$ from dimer $i$ to dimer $j$ on graph $\mathcal{G}_n$ using pCUTs. The reduced amplitudes can again be obtained by subtracting all subcluster contributions of a given graph. Only graphs with up to five links are needed for the calculation of order five; those are also illustrated in Fig.~\ref{fig:graphs_1qp}. The individual hopping amplitudes, embedding factors, and dispersion contributions are given in Appendix~\ref{app:inter_dimer_contributions}.

Upon expressing the dispersion in terms of the structure factor $\gk$ \eqref{gammadef}, we arrive at:
\begin{widetext}
\begin{align}
\frac{\Wk}{J} &=1+d\gk\,\x+ \left\{  \left[ 1+ (-1+\gk) {\lm}^{2
} \right] d-{\frac {{\gk}^{2}}{2}} {d}^{2}\right\} {\x}^{2}+ \left\{
 \left[ {\frac {3\,\gk}{8}}+{\frac {5}{8}}+ \big( -1+\gk
 \big) {\lm}^{2} \right] d \right.\notag\\
&+ \left.\left[ -\gk-{\frac {{\gk}^{2}
}{2}}+ \big( 2\,{\gk}^{2}-2\,\gk \big) {\lm}^{2} \right]
{d}^{2}+{\frac {{\gk}^{3}}{2}}{d}^{3} \right\} {\x}^{3}+ \left\{ \left[ {\frac {11\,\gk}{16}}-{\frac {9}{16}}+ \big( -{\frac {3\,\gamma_{2\vec{k}}}{8}}+{\frac {3\,\gk}{8}} \big) {\lm}^{2}+
 \big( {\frac {3}{4}}-{\frac {3\,\gk}{4}} \big) {\lm}^{4}
 \right] d \right. \notag\\
&+ \left[ -{\frac {15\,{\gk}^{2}}{16}}-{\frac {11\,\gk
}{8}}+{\frac {5}{16}}+ \big( -{\frac {17}{8}}+{\frac {39\,{\gk}^{2
}}{8}}-{\frac {11\,\gk}{4}} \big) {\lm}^{2}+ \big( -{\frac {
1}{4}}-{\frac {\gk}{2}}+{\frac {3\,{\gk}^{2}}{4}} \big) {
\lm}^{4} \right] {d}^{2}  \notag\\
&+ \left. \left[ {\frac {3\,{\gk}^{2}}{2}}+{
\frac {{\gk}^{3}}{2}}+ \big( -{\frac {3\,{\gk}^{2}}{2}}+{\frac
{3\,{\gk}^{3}}{2}} \big) {\lm}^{2} \right] {d}^{3}-{\frac {5
\,{\gk}^{4}}{8}} {d}^{4} \right\} {\x}^{4}\notag\\
&+ \left\{  \left[ -{\frac {
125\,\gk}{128}}-{\frac {3\,\gamma_{2\vec{k}}}{64}}-{\frac {45}{64}}+
 \big( 1-{\frac {17\,\gk}{16}}+{\frac {\gamma_{2\vec{k}}}{16}}
 \big) {\lm}^{2}+ \big( {\frac {5}{4}}-{\frac {9\,\gk}{16}}-
{\frac {11\,\gamma_{2\vec{k}}}{16}} \big) {\lm}^{4} \right] d \right. \notag \\
& + \left[ {\frac {83\,\gk}{64}}-{\frac {15\,{\gk}^{2}}{16}}-{
\frac {17}{64}}+ \big( -{\frac {21}{8}}-{\frac {3\,\gk\,\gamma_{2\vec{k}}}{2}}+{\frac {33\,{\gk}^{2}}{8}} \big) {\lm}^{2}+
 \big( -{\frac {3}{4}}+{\frac {7\,{\gk}^{2}}{8}}-{\frac {\gk}{8
}} \big) {\lm}^{4} \right] {d}^{2} \notag\\
&+  \left[ {\frac {\gk}{32}}+
{\frac {35\,{\gk}^{2}}{16}}+{\frac {33\,{\gk}^{3}}{32}}+ \big(
{\frac {81\,{\gk}^{3}}{8}}-{\frac {45\,\gk}{8}}-{\frac {9\,{
\gk}^{2}}{2}} \big) {\lm}^{2}+ \big( -{\frac {19\,{\gk}^{
2}}{4}}+{\frac {5\,\gk}{8}}+{\frac {33\,{\gk}^{3}}{8}} \big) {
\lm}^{4} \right] {d}^{3} \notag \\
&+ \left.\left[ -{\frac {5\,{\gk}^{3}}{2}}-{
\frac {3\,{\gk}^{4}}{4}}+ \big( {\gk}^{4}-{\gk}^{3} \big)
{\lm}^{2} \right] {d}^{4}+{\frac {7\,{\gk}^{5}}{8}} {d}^{5}
 \right\} {\x}^{5} +\mathcal{O}(\x^6)\,.
\label{wk_smallx}
\end{align}
\end{widetext}
The one-triplon gap is simply obtained as $\Delta=\W_{\vec Q}$. Setting $\lm=0$, one recovers for $d=1$ the known one-triplon gap of the two-leg Heisenberg ladder,\cite{Knetter03_2} while the case $d=2$ reproduces the gap of the square-lattice bilayer.\cite{Weihong97}

\subsection{Double expansion in $K/J$ and $1/d$}

The large-$d$ expansion and the small-$\x$ expansion are expected to match in the double limit
$d\to\infty$ and $\qq\to 0$ -- this is an important cross-check.

First, we re-organize the results of the above expansion in $\x=K/J$, done for arbitrary
$d$, in powers of $\qq=Kd/J$, and extract the leading terms in $1/d$.
For the ground-state energy we find from Eq.~\eqref{e0_smallx}:
\begin{eqnarray}
\label{e0_smallx2}
\frac{E_0}{JN} &=& -\frac{3}{4}
-\frac{3}{8} \frac{\qq^2}{d} +\left(-\frac{3}{16}\,{\qq}^{3}+{\frac {9}{64}}\,{\qq}^{4}\right)\frac{1}{d^2}  \,\nonumber\\
&+&\left({\frac {21}{128}}\,{\qq}^{4}-{\frac {3}{64}}\,{q}^{5}-{\frac {357}{256}}\,{\qq}^{6} - \frac{1}{32}\,\lm^2\qq^6 \right)\frac{1}{d^3}\nonumber\\
&+&\mathcal{O}\left(\frac{1}{d^4}\right).
\end{eqnarray}
Interestingly, the model-dependent selection rules can be used to prove\cite{Coester14} that only terms up to order $\qq^{2m}$ contribute to the $1/d^m$ term in $E_0$, such that the above expression represents the complete expansion up to $1/d^3$ of $E_0$.

For the square of the one-triplon energy we obtain from Eq.~\eqref{wk_smallx}:
\begin{eqnarray}
\frac{\Wk^2}{J^2}&=&1+2\,\gk\,q+\frac{1}{d}\Big\{ \left( 2\,\gk\,{\lm}^{2}-2\,{\lm}^{
2}+2 \right) {\qq}^{2} \nonumber\\
&+& {\phantom{\Big( }} \Big( 6\,{\lm}^{2}{\gk}^{2}-6\,\gk\,{
\lm}^{2}-{\gk}^{2} \Big) {\qq}^{3} \nonumber\\
&+& {\phantom{\Big( }} \Big( 6\,{\lm}^{2}{
\gk}^{3}-6\,{\lm}^{2}{\gk}^{2} \Big) {\qq}^{4}{\phantom{\Big) }}\nonumber\\
&+&{\phantom{\Big( }}\Big( 4\,{
\lm}^{2}{\gk}^{4}-4\,{\lm}^{2}{\gk}^{3} \Big) {\qq}^{5}+\mathcal{O}(\qq^6) \Big\} \nonumber\\
&+&\mathcal{O}\left(\frac{1}{d^2}\right).
\label{wk_smallx2}
\end{eqnarray}
Here, each order in $1/d$ -- with the exception of $d^0$ -- generically receives contributions from all orders in $\qq$.

Second, we expand the results of the $1/d$ expansion from Section~\ref{sec:expa} in $\qq$. Interestingly, the expression for the ground-state energy, Eq.~\eqref{e0_larged}, already has the structure of a small-$\qq$ expansion -- this is related to the model-dependent selection rules as noted above. A comparison of Eq.~\eqref{e0_larged} with Eq.~\eqref{e0_smallx2} shows coincidence.

A look at the large-$d$ triplon dispersion in Eq.~\eqref{Wkexp-a} reveals that the contribution for nonzero $\lm$ does require a small-$\qq$ expansion. Performing this expansion yields exactly the result in Eq.~\eqref{wk_smallx2}.
Hence, the two expansions are found to be consistent in their combined regime of validity, providing an independent check for our $1/d$ expansion results.


\section{Other lattices}
\label{sec:olatt}

So far, we demonstrated the $1/d$ expansion for a simple coupled-dimer model, namely dimers on a hypercubic lattice with nearest-neighbor unfrustrated interactions. More complicated models can be treated, and we give here an incomplete discussion of aspects arising.

\subsection{Interaction terms}

Upon re-writing a general coupled-dimer Heisenberg model \eqref{h} into bond operators, the coefficients of the non-local bilinear, cubic, and quartic terms in the real-space bond-operator Hamiltonian \eqref{hhp} are related to the $K^{mm'}$ in Eq.~\eqref{h} according to
\begin{align}
K_2 &= (K^{11} + K^{22} - K^{12} - K^{21})/2, \notag \\
K_3 &= (K^{11} - K^{22} + K^{12} - K^{21})/2, \notag \\
K_4 &= (K^{11} + K^{22} + K^{12} + K^{21})/2.
\label{k234}
\end{align}
The bilayer model treated in the main part of the paper corresponds to $K_2=K_4=K$ and $K_3=\lm K$.

Eq.~\eqref{k234} shows that the prefactor $K_3$ of the cubic piece vanishes provided that the model remains invariant if in every dimer the spins 1 and 2 are inter-changed (together with all their couplings).
A non-vanishing cubic term occurs if this symmetry is broken, which applies, in addition to the asymmetric bilayer model,\cite{kotov} also, e.g., to the staggered dimer model,\cite{wenzel,fritz11} and to the alternating chain model.\cite{kolezhuk06,zhito06}

Eq.~\eqref{k234} also shows that frustration, introduced by antiferromagnetic $K^{12}$ and $K^{21}$, can induce large quartic couplings which consequently also produce large $1/d$ corrections.

For exchange interactions beyond nearest-neighbor dimers one needs to define a large-$d$ rescaling scheme for every interaction such that a non-trivial large-$d$ limit is obtained. The momentum summations will then involve powers of the different structure factors for the individual interaction terms, and the relevant large-$d$ expansions have to be performed separately for all contributions.

\subsection{Large-$d$ generalizations}

If the $1/d$ expansion is used to access a specific model in $d=2$ or $d=3$ then the first step is a generalization of the model to arbitrary $d$. Depending on the lattice geometry this may be non-trivial, but in any case is {\em not} unique.

Hence, a given model generally admits multiple large-$d$ generalizations which in turn will yield $1/d$ series for observables with different coefficients.
It is interesting to study how the different predictions from low-order $1/d$ expansions differ in such a situation; this will be subject of future work.

\subsection{Finite systems}

The hypercubic-lattice model with linear size $L$ has $N=L^d$ dimer sites, and our results have been derived for the thermodynamic limit where $L\to\infty$ is taken before $d\to\infty$. Interestingly, they also apply to systems where the limit $d\to\infty$ is taken at finite $L$, provided that $L$ is even, as $\sum_{\vec{k}} \gk^2 = N/(2d)$ remains valid.


\section{Summary}
\label{sec:concl}

We have developed a controlled approach to coupled-dimer magnets which can cover the entire phase diagram and smoothly connects the different zero-temperature phases. The method is formulated using bond operators and utilizes $1/d$ as a small parameter, i.e., provides systematic $1/d$ expansions for any observable. Phrased differently, we have identified a small parameter -- $1/d$ -- which controls the well-known bond-operator approach and utilizes a systematic improvement of the frequently used leading-order calculations.

In this paper, we have demonstrated the method for a $d$-dimensional hypercubic-lattice generalization of the bilayer Heisenberg model and determined the ground-state energy, the one-triplon spectrum and weight, and the phase boundary to the antiferromagnetic phase.

Contact was made with a perturbative expansion in the inter-dimer coupling, performed using continuous unitary transformation. High-order results of this expansion were used as a cross-check of the $1/d$ expansion.

In the companion paper Ref.~\onlinecite{ii}, the $1/d$ expansion is applied to the antiferromagnetically ordered phase of the hypercubic coupled-dimer model, where it is shown that both phases can be smoothly connected order by order in $1/d$, as expected on general grounds.

We envision further applications of the $1/d$ expansion developed here to coupled-dimer magnets in a uniform field, where Bose-Einstein condensation of triplons occurs\cite{gia08} and to systems with geometric frustration,\cite{bamno,bacro} including cases with non-collinear and incommensurate order. The harmonic bond-operator approach has been applied to systems with quenched disorder,\cite{mv13} and we expect insights into corrections here as well.


\acknowledgments

We thank E. Andrade, S. Burdin, F. H. L. Essler, D. K. Morr, G. S. Uhrig, and M. E. Zhitomirsky for helpful discussions.
This research has been supported by the DFG (GRK 1621 and SFB 1143), the GIF (G 1025-36.14/2009), and by the Virtual Institute VI-521 of the Helmholtz association.


\appendix

\section{Projectors and spin commutation relations}
\label{app:commu}

Here we discuss the choice of projection operators $P_i$ used to express the spin operators in
terms of hard-core triplet operators as in Eq.~\eqref{sproj}.
Under the assumption that $P_i$ is an arbitrary function of $n_i=\sum_\gamma t_{i\gamma}^\dagger
t_{i\gamma}$, an explicit computation yields
\begin{align}
[S^\alpha_{im}, S^\beta_{im'}]_- = \ii \epsilon_{\alpha\beta\gamma}
& S^\gamma_{im} \delta_{mm'} \notag\\
+ \frac{(-1)^{m+m'}}{4} [&t ^{\dagger}_{i\alpha} (P^2_i-1) t_{i\beta} - P_i t ^{\dagger}_{i\alpha} t_{i\beta}
P_i \notag \\
- &t^{\dagger}_{i\beta} (P^2_i-1) t_{i\alpha} + P_i t^{\dagger}_{i\beta}t_{i\alpha} P_i ].
\label{comgen}
\end{align}
The first line corresponds to the standard spin commutator, and the extra terms can be written, using
$P_i = f(n_i)$, as
\begin{equation}
(t^{\dagger}_{i\alpha} t_{i\beta} - t^{\dagger}_{i\beta} t_{i\alpha})
\left[f^2(n_i-1) - 1 - f^2(n_i)\right].
\label{comex}
\end{equation}

Given the representation \eqref{sproj} of the spin operators, we have to require $f(0)=1$ and $f(1)=0$,
such that any matrix element of $\vec{S}_{im}$ between physical and unphysical states is suppressed.
With this requirement, the extra terms \eqref{comex} automatically vanish within the physical Hilbert
space defined by $n_i \leq 1$. Hence, at a formal level, the choice of projection operator is {\em not}
unique, i.e., any function with $f(0)=1$ and $f(1)=0$ could be chosen.

However, for practical purposes, $f(x)=1-x$ as in Eq.~\eqref{linproj} is most efficient, because a
non-linear function would lead to a more complicated Hamiltonian with a more involved normal-ordering
procedure.

Non-analytic choices of $f(x)$ may lead to even more severe problems: It is illuminating to consider
the choice $f(x) = \sqrt{1-x}$ which might have appeared suitable based on similarities to the
Holstein-Primakoff representation of spin operators in the context of spin-wave theory:\cite{hp1940}
There, the Hilbert-space constraint for Holstein-Primakoff bosons, $n\leq 2S$, is implemented via
square-root projectors, $S^- = a^\dagger \sqrt{2S - a^\dagger a}$, $S^+ = \sqrt{2S - a^\dagger a}\, a$.
An important difference, however, is that in spin-wave theory (when used as an asymptotic expansion)
the physical Hilbert space is infinite at the reference point $S\to\infty$, such that matrix elements
with unphysical states formally do not appear. In our case, the physical Hilbert space is finite, such
that $P_i$ practically has to be evaluated also with $n_i \geq 2$ states. Now, $P_i$ is defined via the
series expansion of $f(n_i)$, and the series of $\sqrt{1-x}$ is non-convergent for $x>1$ -- this
renders calculations with square-root projectors impossible.


\section{Momentum sums in large $d$ and expectation values}
\label{app:expval}

As already mentioned in Section~\ref{sec:ham}, the basis for the $1/d$ expansion is the observation that
the magnitude of $\gk$ \eqref{gammadef} scales as $1/\sqrt{d}$ for typical $\vec k$ and large $d$. This
implies that, inside a $\vec k$ summation, $\gk$ can be treated as a small parameter, and a formal
$1/d$ expansion can be generated by expanding in $\gk$. Direct summations over $\gk$, using $\int_0^{2\pi} dx \cos^2 x = 1/2$ etc., yield:
\begin{align}
\label{sum1}
&\frac{1}{N} \sum_{\vec k} \gk^{2n+1} = 0\,,\\
&\frac{1}{N} \sum_{\vec k} \gk^2 = \frac{1}{2d}\,,~~
\frac{1}{N} \sum_{\vec k} \gk^4 = \frac{3}{4d^2} - \frac{3}{8d^3}\,, \\
\label{sum2}
&\frac{1}{N} \sum_{\vec k} \gamma_{\vec k+ \vec k''} \gamma_{\vec k + \vec k'} = \frac{\gamma_{\vec k'-\vec k''}}{2d}\,.
\end{align}
The expressions for physical observables arise from the bilinear Hamiltonian \eqref{hh2k} and its
perturbations and involve combinations of the mode energy $\wk$ \eqref{om0} and Bogoliubov coefficients \eqref{eq3}. For the $1/d$ expansion these need to be expanded in $\gk$:
\begin{align}
\frac{\wk}{J} &= 1+\gk\qq - \frac{\gk^2\qq^2}{2} + \frac{\gk^3\qq^3}{2} - \frac{5\gk^4\qq^4}{8} + \mathcal{O}(\gk^5) \,,\notag\\
v_{\vec{k}}^{2} &= \frac{\gk^{2} q^{2}}{4} - \frac{\gk^{3} q^{3}}{2} + \frac{15 \gk^{4} q^{4}}{16} + \mathcal{O}(\gk^5), \notag\\
u_{\vec{k}}^{2} &= 1 + v_{\vec{k}}^{2}, \notag\\
u_{\vec{k}}v_{\vec{k}} &= -\frac{\gk q}{2} + \frac{\gk^{2} q^{2}}{2} - \frac{3\gk^{3} q^{3}}{4} + \frac{5\gk^{4} q^{4}}{4} + \mathcal{O}(\gk^5) \,.
\label{expingak}
\end{align}
Frequently needed are the momentum sums defined in Eq.~\eqref{rdef}. Using Eqs.~\eqref{sum1}, \eqref{sum2}, and \eqref{expingak}, their large-$d$ expansion is found as follows:
\begin{align}
\label{r1}
R_1 &= \frac{1}{N} \sum_{\vec k} u_{\vec k} v_{\vec k} ~~~= \frac{q^2}{4d} + \frac{15 q^{4}}{16 d^2} + \mathcal{O}(d^{-3})
\end{align}
\begin{align}
\label{r2}
R_2 &= \frac{1}{N} \sum_{\vec k} v^2_{\vec k} ~~~~~~= \frac{q^2}{8d} + \frac{45 q^{4}}{64 d^2} + \mathcal{O}(d^{-3})
\\
\label{r3}
R_3 &= \frac{1}{N} \sum_{\vec k} \gk u_{\vec k} v_{\vec k} = -\frac{q}{4d} - \frac{9 q^{3}}{16 d^2} + \mathcal{O}(d^{-3}) \\
\label{r4}
R_4 &= \frac{1}{N} \sum_{\vec k} \gamma_{\vec k} v^2_{\vec k} ~~~\,= -\frac{3 q^{3}}{8 d^2} + \mathcal{O}(d^{-3})
\end{align}
The $R_{1\ldots4}$ are related to expectation values of the bilinear Hamiltonian \eqref{hh2k} as
follows:
\begin{align}
\sum_{i} \langle t_{i\alpha}^\dagger t_{i\beta}^\dagger \rangle &= N \delta_{\alpha\beta} R_1 \,,~
\sum_{i} \langle t_{i\alpha}^\dagger t_{i\beta} \rangle = N \delta_{\alpha\beta} R_2, \notag \\
\sum_{\langle ij \rangle} \langle t_{i\alpha}^\dagger t_{j\beta}^\dagger \rangle &= Nd \delta_{\alpha\beta} R_3\,,
\sum_{\langle ij \rangle} \langle t_{i\alpha}^\dagger t_{j\beta} \rangle = Nd \delta_{\alpha\beta} R_4\,.
\end{align}
Note, however, that the full $1/d$ expansion for these expectation values also involve corrections
from the additional Hamiltonian pieces $\mathcal{H}_{2b,4,6}$. These corrections ensure that
$\langle t_{i\alpha}^\dagger t_{i\alpha}^\dagger \rangle = 0$ order by order in the $1/d$ expansion, as required by the hard-core constraint, see Section~\ref{sec:expalocal}.

Finally, we also need the following higher-order combination of Bogoliubov coefficients:
\begin{align}
\label{r5}
R'_5(\vec{k}) &= \frac{1}{N} \sum_{\vec k'} u_{\vec k'} v_{\vec k'} u_{\vec{k}-\vec{k'}} v_{\vec{k}-\vec{k'}} = \frac{\gk\qq^2}{8d}\,.
\end{align}


\section{Cubic and quartic vertex functions}

\label{app:vertices}

The cubic vertex functions are linearly proportional to the asymmetry parameter $\lm$ \eqref{lamdef} and read:
\begin{align}
\vca(123) &= -\ii \lm \qq J \gamma_{2+3} (u_1 u_2 v_3 - v_1 v_2 u_3), \\
\vcb(123) &= -\ii \lm \qq J \gamma_{2-3} (u_1 u_2 u_3 - v_1 v_2 v_3), \\
\vcc(123) &= -\ii \lm \qq J \gamma_{2+3} (v_1 u_2 v_3 - u_1 v_2 u_3), \\
\vcd(123) &= -\ii \lm \qq J \gamma_{2-3} (u_1 v_2 v_3 - v_1 u_2 u_3).
\end{align}

The quartic vertex functions are:
\begin{widetext}
\begin{align}
\vqa(1234) &= \frac{\qq J}{2} \sum_{\alpha, \beta,\alpha \neq \beta}
(\gamma_{2+3} u_1 v_2 u_3 v_4 - \gamma_{2+4} u_1 u_2 v_3 v_4) \nonumber \\
&- \qq J \sum_{\alpha,\beta} (\gamma_{2} u_1 v_2 u_3 v_4 + \gamma_{2} u_1 u_2 u_3 v_4 + \gamma_{2} v_1 v_2 v_3 u_4 + \gamma_{2+3+4} u_1 v_2 u_3 v_4),
\end{align}
\begin{align}
\vqb(1234) &= \frac{\qq J}{2} \sum_{\alpha, \beta,\alpha \neq \beta}
(\gamma_{2-4} u_1 v_2 u_3 v_4 + \gamma_{2-4} v_1 u_2 v_3 u_4 - \gamma_{2-4} u_1 u_2 u_3 u_4 - \gamma_{2-4} v_1 v_2 v_3 v_4) \nonumber \\
&- \qq J \sum_{\alpha,\beta} (\gamma_{2} u_1 v_2 u_3 v_4 + \gamma_{4} v_1 u_2 v_3 u_4 + \gamma_{2-3-4} u_1 v_2 u_3 v_4
+ \gamma_{1+2-4} v_1 u_2 v_3 u_4 + \gamma_{2} u_1 u_2 v_3 u_4 + \gamma_{4} u_1 v_2 v_3 v_4 \nonumber \\
&~~~~~~~~~~~~+ \gamma_{3} u_1 v_2 u_3 u_4 + \gamma_{1} v_1 v_2 v_3 u_4),
\\
\vqc(1234) &= \frac{\qq J}{2} \sum_{\alpha, \beta,\alpha \neq \beta}
(\gamma_{2-3} u_1 u_2 u_3 u_4 + \gamma_{-3-4} u_1 v_2 u_3 v_4 + \gamma_{1+2} v_1 u_2 v_3 u_4 + \gamma_{1-4} v_1 v_2 v_3 v_4 \nonumber \\
&~~~~~~~~~~~~~~~~~-\gamma_{2-3} u_1 v_2 v_3 u_4 - \gamma_{-3-4} u_1 v_2 v_3 u_4 - \gamma_{1+2} u_1 v_2 v_3 u_4 - \gamma_{1-4} u_1 v_2 v_3 u_4) \nonumber \\
&- \qq J \sum_{\alpha,\beta}
(\gamma_{3} u_1 u_2 u_3 u_4 + \gamma_{3} u_1 v_2 u_3 v_4 + \gamma_{1} v_1 u_2 v_3 u_4 + \gamma_{1} v_1 v_2 v_3 v_4
+\gamma_{2-3-4} u_1 u_2 u_3 u_4 + \gamma_{2-3-4} u_1 v_2 u_3 v_4 \nonumber \\
&~~~~~~~~~~~~+ \gamma_{1+2-4} v_1 u_2 v_3 u_4 + \gamma_{1+2-4} v_1 v_2 v_3 v_4
+\gamma_{3} u_1 u_2 v_3 u_4 + \gamma_{3} u_1 v_2 v_3 v_4 + \gamma_{1} u_1 u_2 v_3 u_4 + \gamma_{1} u_1 v_2 v_3 v_4 \nonumber \\
&~~~~~~~~~~~~+\gamma_{1} v_1 u_2 u_3 u_4 + \gamma_{1} v_1 v_2 u_3 v_4 + \gamma_{3} v_1 u_2 u_3 u_4 + \gamma_{3} v_1 v_2 u_3 v_4),
\end{align}
\begin{align}
\vqd(1234) &= \frac{\qq J}{2} \sum_{\alpha, \beta,\alpha \neq \beta}
(\gamma_{2+3} u_1 v_2 u_3 u_4 + \gamma_{2-4} u_1 v_2 v_3 v_4 + \gamma_{1-4} u_1 v_2 u_3 u_4 + \gamma_{1+3} u_1 v_2 v_3 v_4 \nonumber \\
&~~~~~~~~~~~~~~~~~-\gamma_{2+3} u_1 u_2 v_3 u_4 - \gamma_{2-4} u_1 u_2 v_3 u_4 - \gamma_{2-4} v_1 v_2 u_3 v_4 - \gamma_{2+3} v_1 v_2 u_3 v_4) \nonumber \\
&- \qq J \sum_{\alpha,\beta}
(\gamma_{2} u_1 v_2 u_3 u_4 + \gamma_{2} u_1 v_2 v_3 v_4 + \gamma_{4} u_1 v_2 u_3 u_4 + \gamma_{3} u_1 v_2 v_3 v_4
+\gamma_{2+3-4} u_1 v_2 u_3 u_4 + \gamma_{2+3-4} u_1 v_2 v_3 v_4 \nonumber \\
&~~~~~~~~~~~~+ \gamma_{1+2-4} u_1 v_2 u_3 u_4 + \gamma_{1+2+3} u_1 v_2 v_3 v_4
+\gamma_{2} u_1 u_2 u_3 u_4 + \gamma_{2} u_1 u_2 v_3 v_4 + \gamma_{4} u_1 v_2 u_3 v_4 + \gamma_{3} u_1 v_2 u_3 v_4 \nonumber \\
&~~~~~~~~~~~~+\gamma_{3} u_1 v_2 v_3 u_4 + \gamma_{4} u_1 v_2 v_3 u_4 + \gamma_{2} v_1 v_2 u_3 u_4 + \gamma_{2} v_1 v_2 v_3 v_4 ).
\end{align}
\end{widetext}


\section{Evaluation of diagrams in a $1/d$ expansion}
\label{app:dgrdemo}

Here we demonstrate the evaluation of Feynman diagrams in a $1/d$ expansion, using a sample self-energy diagram involving two cubic vertices, with the full structure of the cubic Hamiltonian piece given in Eq.~\eqref{hp3}.
To be explicit, we focus on a normal self-energy diagram with two $\vca$ vertices which furthermore have $\tau_x$ as external legs with frequency $\w$ and momentum $\vec{k}$, Fig.~\ref{dgrdemo}. Its explicit expression reads:
\begin{align}
\label{se}
 \Sigma_{\Gamma} &=  \frac{i}{2\pi} \int d\w_1 d\w_2 \frac{1}{N}\sum_{{\vec k}_1{\vec k}_2} \Gamma(\vec{k},\vec{k}_1,\vec{k}_2)
 \mathcal{G}_{0N}(\vec{k}_1,\w_1) \nonumber\\ &~~~~\times \mathcal{G}_{0N}(\vec{k}_2,\w_2) \delta(\w+\w_1+\w_2) \delta_{\vec{k}+\vec{k}_1+\vec{k}_2}
\end{align}
where $\mathcal{G}_{0N}$ is the normal $\tau$ Green's function for the unperturbed Hamiltonian,
\begin{equation}
\mathcal{G}_{0N}(\vec{k},\omega) = \frac{1}{\omega - \omega_k}
\end{equation}
and $\Gamma(\vec{k},\vec{k}_1,\vec{k}_2)$ represents the product of vertex functions and respective permutations of the legs of the cubic vertex corresponding to this diagram
($\vec{k}_y \equiv \vec{k}_1$, $\vec{k}_z \equiv \vec{k}_2, \w_{\vec{k}_y} \equiv \w_1, \w_{\vec{k}_z} \equiv \w_2$), i.e.:

\begin{figure}[!t]
\includegraphics[width=0.47\textwidth]{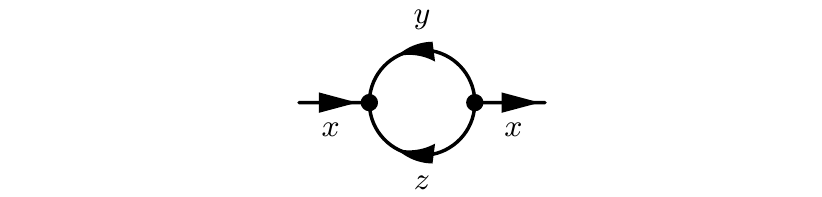}
\caption{Sample self-energy diagram with two cubic $\vca$ vertices.}
\label{dgrdemo}
\end{figure}

\begin{align}
&\Gamma(\vec{k},\vec{k}_1,\vec{k}_2) = \Gamma_1(\vec{k},\vec{k}_1,\vec{k}_2) + \Gamma_2(\vec{k},\vec{k}_1,\vec{k}_2) + \Gamma_3(\vec{k},\vec{k}_1,\vec{k}_2)
\end{align}
with
\begin{align}
&\Gamma_1 (\vec{k},\vec{k}_1,\vec{k}_2)= \nonumber \\
&2\left[ \vca(\vec{k}\vec{k}_1\vec{k}_2)\vca(\vec{k}\vec{k}_1\vec{k}_2)-\vca(\vec{k}\vec{k}_1\vec{k}_2)\vca(\vec{k}\vec{k}_2\vec{k}_1)  \right. \nonumber \\
&+\vca(\vec{k}\vec{k}_1\vec{k}_2)\vca(\vec{k}_2\vec{k}\vec{k}_1)-\vca(\vec{k}\vec{k}_1\vec{k}_2)\vca(\vec{k}_1\vec{k}\vec{k}_2)   \nonumber \\
&\left.+\vca(\vec{k}\vec{k}_1\vec{k}_2)\vca(\vec{k}_1\vec{k}_2\vec{k})-\vca(\vec{k}\vec{k}_1\vec{k}_2)\vca(\vec{k}_2\vec{k}_1\vec{k})\right], \\
&\Gamma_2 (\vec{k},\vec{k}_1,\vec{k}_2)=  \nonumber \\
&2\left[ \vca(\vec{k}_2\vec{k}\vec{k}_1)\vca(\vec{k}\vec{k}_1\vec{k}_2)-\vca(\vec{k}_2\vec{k}\vec{k}_1)\vca(\vec{k}\vec{k}_2\vec{k}_1)  \right. \nonumber \\
&+\vca(\vec{k}_2\vec{k}\vec{k}_1)\vca(\vec{k}_2\vec{k}\vec{k}_1)-\vca(\vec{k}_2\vec{k}\vec{k}_1)\vca(\vec{k}_1\vec{k}\vec{k}_2)   \nonumber \\
&\left.+\vca(\vec{k}_2\vec{k}\vec{k}_1)\vca(\vec{k}_1\vec{k}_2\vec{k})-\vca(\vec{k}_2\vec{k}\vec{k}_1)\vca(\vec{k}_2\vec{k}_1\vec{k})\right],
\end{align}
\begin{align}
&\Gamma_3 (\vec{k},\vec{k}_1,\vec{k}_2)=  \nonumber \\
&2\left[ \vca(\vec{k}_1\vec{k}_2\vec{k})\vca(\vec{k}\vec{k}_1\vec{k}_2)-\vca(\vec{k}_1\vec{k}_2\vec{k})\vca(\vec{k}\vec{k}_2\vec{k}_1)  \right. \nonumber \\
&+\vca(\vec{k}_1\vec{k}_2\vec{k})\vca(\vec{k}_2\vec{k}\vec{k}_1)-\vca(\vec{k}_1\vec{k}_2\vec{k})\vca(\vec{k}_1\vec{k}\vec{k}_2)   \nonumber \\
&\left.+\vca(\vec{k}_1\vec{k}_2\vec{k})\vca(\vec{k}_1\vec{k}_2\vec{k})-\vca(\vec{k}_1\vec{k}_2\vec{k})\vca(\vec{k}_2\vec{k}_1\vec{k})\right].
\end{align}
The factors of $2$ arise from permutations that yield identical contributions as the ones that appear above, e.g.,
$\vca(\vec{k}\vec{k}_2\vec{k}_1)\vca(\vec{k}\vec{k}_2\vec{k}_1) \hat{=} \vca(\vec{k}\vec{k}_1\vec{k}_2)\vca(\vec{k}\vec{k}_1\vec{k}_2)$.

For the purpose of illustration we will now show the explicit calculation for $\Gamma_1$.
We first perform the frequency integral in \eqref{se}.
The resulting expression is then 
\begin{equation}
\label{sem}
\Sigma_{\Gamma_1} = \frac{1}{N} \sum_{{\vec k}_1{\vec k}_2} \frac{\Gamma_1(\vec{k},\vec{k}_1,\vec{k}_2)}{-\w-\w_1-\w_2} \, \delta_{\vec{k}+\vec{k}_1+\vec{k}_2}\,.
\end{equation}
The remaining momentum integration is the central element of the $1/d$ expansion. We recall that momentum sums of various powers of $\gk$ scale as powers of $1/d$, see equations \eqref{sum1} and \eqref{sum2}. In particular, a momentum sum of $\gk^{2n+1}$ is zero and that of $\gk^{2n}$ scales as $1/d^n$ (plus higher-order terms). As a consequence, any function $f(\gk)$ under a momentum integral can be Taylor-expanded in $\gk$ as to generate an expansion in $1/d$ {\em after} the momentum integration. (Note that our small control parameter is $1/d$, not $\gk$.)

The actual calculation requires the $\gk$ expansions of the mode energy and the Bogoliubov coefficients, Eq.~\eqref{expingak}, as input. As we restrict our attention to the leading $1/d$ order of the self-energy, we can approximate $u_{\vec{k}}^{2} \approx 1$ and $\wk \approx J$,
since since $\Gamma_1$ involves factors of $\gk$ and $v_{\vec{k}}$ which will generate at least one factor of $1/d$. (Obtaining higher orders is straightforward, but tedious, and requires to include higher orders for $u_{\vec{k}}^{2}$ and $\wk$.) Hence, to order $1/d$ Eq.~\eqref{sem}
reduces to
\begin{equation}
\label{sem1}
\Sigma_{\Gamma_1} = -\frac{1}{\w+2J} \frac{1}{N}\sum_{\vec{k}_1}  \Gamma_1(\vec{k},\vec{k}_1,-\vec{k}_1-\vec{k})\,.
\end{equation}
Here we now need to collect those terms which are $\mathcal{O}(\gk^{2})$, as
$\sum \gk^{2} \propto 1/d$ -- these are terms like $u_{\vec{k}}^{2}u_{\vec{k}_1}^{2}v_{-\vec{k}-\vec{k}_1}^{2}$ etc. This yields
\begin{widetext}
\begin{align}
\label{sem2}
\Sigma_{\Gamma_1} &= -\frac{2 \gk \lm^{2} K^{2}}{\w+2J} \frac{1}{N}\sum_{\vec{k}_1} \left[  \gk u_{\vec{k}}^{2} u_{\vec{k}_1}^{2} v_{-\vec{k}-\vec{k}_1}^{2}  \right.
-2\gk u_{\vec{k}}v_{\vec{k}} u_{\vec{k}_1}v_{\vec{k}_1} u_{-\vec{k}-\vec{k}_1}v_{-\vec{k}-\vec{k}_1} + \gk v_{\vec{k}}^{2} v_{\vec{k}_1}^{2} u_{-\vec{k}-\vec{k}_1}^{2} \nonumber \\
&-\gk u_{\vec{k}}^{2} u_{\vec{k}_1}v_{\vec{k}_1} u_{-\vec{k}-\vec{k}_1}v_{-\vec{k}-\vec{k}_1}  +2\gk u_{\vec{k}}v_{\vec{k}} u_{\vec{k}_1}^{2} v_{-\vec{k}-\vec{k}_1}^{2}
-\gk v_{\vec{k}}^{2} u_{\vec{k}_1}v_{\vec{k}_1} u_{-\vec{k}-\vec{k}_1}v_{-\vec{k}-\vec{k}_1}
+ u_{\vec{k}}v_{\vec{k}} \gamma_{\vec{k}_1} u_{\vec{k}_1}^{2} u_{-\vec{k}-\vec{k}_1}v_{-\vec{k}-\vec{k}_1} \nonumber \\
&-v_{\vec{k}}^{2} \gamma_{\vec{k}_1} u_{\vec{k}_1}v_{\vec{k}_1} v_{-\vec{k}-\vec{k}_1}^{2} - u_{\vec{k}}v_{\vec{k}} u_{\vec{k}_1}^{2} \gamma_{\vec{k}_2} u_{-\vec{k}-\vec{k}_1}v_{-\vec{k}-\vec{k}_1}
\left. +v_{\vec{k}}^{2} u_{\vec{k}_1}v_{\vec{k}_1} \gamma_{\vec{k}_2} u_{-\vec{k}-\vec{k}_1}^{2}\right]\,.
\end{align}
Using the definitions of $R_{1\ldots5}$ in Eqs. \eqref{rdef} and \eqref{rdef1}, this can be converted into
\begin{align}
\label{sem3}
\Sigma_{\Gamma_1} = -\frac{2 \gk \lm^{2} K^{2}}{\w+2J} \left\{ \gk u_{\vec{k}}^{2} \big[ R_2 - R'_5(\vec{k})\big]
+ v_{\vec{k}}^{2} \big[\gk R_2 - \gk R'_5(\vec{k}) -R_3 + R'_3(\vec{k})\big]
+ u_{\vec{k}}v_{\vec{k}} \big[2\gk R_2 - 2\gk R'_5(\vec{k}) -R_3 + R'_3(\vec{k})\big]\right\}.
\end{align}
A similar calculation for the $\Gamma_2$ and $\Gamma_3$ combination of vertices results in:
\begin{align}
\Sigma_{\Gamma_2} &= \mathcal{O}(1/d^{2}), \\
\Sigma_{\Gamma_3} &= -\frac{2  \lm^{2} K^{2}}{\w+2J} \left\{v_{\vec{k}}^{2} \left[ \gk R'_3(\vec{k}) -\gk R_3 + \frac{1}{2d} - \frac{\gk}{2d}\right]
+ \gk u_{\vec{k}}v_{\vec{k}} \left[ R'_3(\vec{k}) - R_3\right]\right\}.
\end{align}
\end{widetext}
Summing $\Sigma^{4(b)}=\Sigma_{\Gamma_1}+\Sigma_{\Gamma_2}+\Sigma_{\Gamma_3}$ finally yields Eq.~\eqref{cubic2} of the main text.

The calculation for other diagrams used in this paper follows the same strategy as outlined here; typically only a small fraction of possible vertex contributions of a given diagram eventually contributes to order $\mathcal{O}(1/d)$. For higher orders, the use of computer algebra is indispensable.


\section{Brueckner approach}
\label{app:kotov}

In this Appendix, we discuss the possibility to generate a $1/d$ expansion using an different treatment of the hard-core constraint \eqref{hardcore} of the triplet excitations: Instead of the projectors \eqref{linproj}, the infinite on-site repulsion $\mathcal{H}_U$ \eqref{HU} is treated using the Brueckner approach as proposed in Ref.~\onlinecite{kotov}. This approximation is known to be controlled in the dilute-gas limit, and since we know that the triplet density scales as $1/d$ we expect that the Brueckner approximation becomes accurate here as well.

Following Ref.~\onlinecite{kotov}, we introduce a renormalized quartic vertex, resulting from the hard-core repulsion $\mathcal{H}_U$ \eqref{HU} of the $t$ particles, which is obtained from a self-consistent ladder summation:\cite{kotov}
\begin{equation}
\label{Gamma}
\Gamma({\vec k},\omega)=
-\left(\frac{1}{N}\sum_{\vec p}\frac{
u_{{\vec p}}^2 u_{{\vec k-\vec p}}^2}
{\omega-\omega_{\vec p}-\omega_{\vec k- \vec p}}\right)^{-1}.
\end{equation}
Here, all anomalous scattering vertices have been neglected, which is justified
in the small-density limit.

We now proceed to calculate corrections to the triplon dispersion, both from $\mathcal{H}_U$ and from the quartic terms in the Hamiltonian (note that these are only the terms in the first line of Eq.~\eqref{hh4}, while those in the other lines arise from the projectors and are absent here). For simplicity, we restrict ourselves to the symmetric case $\lm=0$.
Importantly, the diagrammatics is done here directly for the $t$ particles, i.e., the following self-energies and propagators are those of $t$ particles.

To leading order, the normal self-energy from ${\cal H}_U$ is given by the sum of Hartree and Fock diagrams:
\begin{eqnarray}
\label{Sigma_U}
\Sigma_\alpha^U({\vec k},\omega)&=&
\Sigma_{\alpha\alpha}({\vec k},\omega) +
\sum_\beta \Sigma_{\alpha\beta} ({\vec k},\omega)\,, \nonumber\\
\Sigma_{\alpha\beta} ({\vec k},\omega) &=&
\frac{1}{N}\sum_{\vec q}
v_{{\vec q}\beta}^2\Gamma_{\alpha\beta,\alpha\beta}({\vec k + \vec q},\omega-\omega_{{\vec q}\beta}).
\end{eqnarray}
Spin indices $\alpha,\beta$ are written here for book-keeping purposes only; both the $\Gamma$ vertex and the self-energies do not depend on $\alpha,\beta$ in the paramagnetic phase.

Anticipating that $\Sigma_\alpha^U \propto 1/d$ we conclude that the renormalized vertex $\Gamma$ will be of order $1/d^0$. Restricting our attention to this leading contribution, we can approximate  $u^{2}_{\vec k} = 1$, $v^{2}_{\vec k} = 0$, and the mode energy $\w_{\vec p}=J$, such that Eq.~\eqref{Gamma} immediately gives
\begin{equation}
\label{gamma2}
\Gamma({\vec k},\omega)= - (\omega - 2J) + \mathcal{O}\left(\frac{1}{d}\right)\,.
\end{equation}
The normal self-energy from Eq.~\eqref{Sigma_U} then evaluates to
\begin{equation}
\Sigma_{N}^{U}({\vec k},\omega) = - \frac{\qq^{2} (\omega - 3J)}{2d}
\end{equation}
up to order $1/d$, where we have again set the mode energy to $J$ and used the momentum-summation result \eqref{r2}.
In addition, there is an anomalous self-energy contribution from the $\Gamma$ vertex.\cite{Kotov_new} This is expressed as follows:
\begin{equation}
\Sigma_{A}^{U}({\vec k},\omega)= \frac{1}{N}\sum_{\vec p} u_{\vec p} v_{\vec p} \Gamma(0,0)
\end{equation}
Evaluating this in the large-$d$ limit as before, and using Eq.~\eqref{r1}, we find:
\begin{equation}
\Sigma_{A}^{K}({\vec k},\omega)= \frac{1}{N}\sum_{\vec p} u_{\vec p} v_{\vec p} 2 J =  J \frac{\qq^{2}}{2 d}\,.
\end{equation}
Finally, we take into account the quartic terms of the triplon Hamiltonian (not arising from projectors) -- this is done on the Hartree-Fock level \cite{kotov}. Summing all self-energy contributions we have:
\begin{align}
\label{sigman}
\Sigma_{N}({\vec k},\omega) &= -\frac{\qq^{2} (\omega - 3J)}{2 d} + 2 \gamma_{\vec k} J R_4, \\
\Sigma_{A}({\vec k},\omega) &= J \frac{\qq^{2}}{2 d} - 2 \gamma_{\vec k} J R_3.
\label{sigmaa}
\end{align}

These self-energies enter the Dyson equation for the $t$ particles from which we want to extract the renormalized mode energy to order $1/d$. One difference to the Dyson equation for the $\tau$ particles used in the main paper is that the anomalous self-energy cannot be neglected here, as the $t$ particles display anomalous propagators already to leading order.
The Dyson equation can be cast in the following form
\begin{widetext}
\begin{equation}
\label{gtbrueckner}
\mathcal{G}(\vec k, \omega) = \frac{\omega Z^{-1} + A_{\vec k} + \Sigma_{N}(\vec k,0)}{\omega^{2} \big(Z^{-2} - \Sigma'^2_{A}(\vec k,0)\big) - \big(A_{\vec k} + \Sigma_N(\vec k,0)\big)^{2}
+ \big(B_{k} + \Sigma_{A}(\vec k,0)\big)^{2}}
\end{equation}
\end{widetext}
where
\begin{align}
Z^{-1} &= 1 - \frac{\partial \Sigma_{N}(\vec k, \omega)}{\partial \omega} |_{\omega = 0} = 1 + \frac{\qq^{2}}{2 d}, \\
\Sigma'_{A}(\vec k,0) &=  \frac{\partial \Sigma_{A}(\vec k, \omega)}{\partial \omega} |_{\omega = 0} = 0;
\end{align}
note that an expansion of $\Sigma(\vec k,\w)$ (\ref{sigman},\ref{sigmaa}) around $\w=0$ is exact here.
The pole of the $t$ Green's function \eqref{gtbrueckner} follows the equation:
\begin{equation}
\W_k^{2} = \frac{\big(A_{k} + \Sigma_{N}(\vec k,0)\big)^{2} - \big(B_{k} + \Sigma_{A}(\vec k,0)\big)^{2}}{Z^{-2} - \Sigma'^2_{A}(\vec k,0)} \,.
\end{equation}
Thus we obtain the expansion of the dispersion relation to order $1/d$,
\begin{equation}
\frac{\W_k^{2}}{J^2} = 1 + 2\gamma_{\vec k} \qq + \frac{1}{d} (2 \qq^{2} - \gamma_{\vec k}^{2} \qq^{3})\,,
\end{equation}
identical to the result \eqref{Wkexp} derived in the body of the paper.

This indicates that the Brueckner approximation is indeed controlled in the large-$d$ limit. We note, however, that it cannot be easily used to systematically generate higher orders of the $1/d$ expansion, first, because the $\Gamma$ vertex becomes extremely complicated beyond leading order, and second, because a fully consistent diagrammatic treatment needs to be formulated in $\tau$ instead of $t$ particles to ensure Wick's theorem.



\section{Specific contributions to the inter-dimer perturbation theory}
\label{app:inter_dimer_contributions}

In this Appendix we give more details for the calculation of the one-triplon dispersion up to order 5 perturbation theory in $\x$. The dispersion is obtained by Fourier transformation of the reduced one-triplon hopping amplitudes. The non-vanishing reduced hopping amplitudes up to order 5, all given in units of $J$, read:
\begin{eqnarray}
 t^{(0)}_{0,0} &=& 1,\\
 t^{(1)}_{0,1} &=& \frac{1}{2}\,\x+\frac{1}{2}\,{\lm}^{2}\x^2-\frac{1}{2}\,{\lm}^{4}\x^4,\\
 t^{(1)}_{0,0} &=& -\frac{1}{2}\,{\lm}^{2}\x^2+\frac{1}{2}\,{\lm}^{4}\x^4+\frac{3}{8}\,{\x}^{2}+\frac{3}{16}\,{\x}^{3}\nonumber\\
             &&+{\frac{3}{128}}\,{\x}^{4}-{\frac {15}{256}}\,{\x}^{5},\\
 t^{(2)}_{0,0} &=& -{\frac{11}{128}}\,{\x}^{4}-{\frac{85}{512}}\,{\x}^{5}-\frac{1}{16}\,{\lm}^{2}{\x}^{4}-{\frac{139}{384}}\,{\lm}^{2}{\x}^{5}\nonumber\\
             &&-\frac{1}{8}\,{\lm}^{4}\x^4-{\frac{7}{16}}\,{\lm}^{4}\x^5,\\
 t^{(2)}_{0,1} &=& -\frac{1}{16}\,{\x}^{3}-{\frac{5}{64}}\,{\x}^{4}-{\frac{31}{512}}\,{\x}^{5}-\frac{1}{4}
\,\lm^2\x^3-{\frac{5}{16}}\,\lm^2{\x}^{4}\nonumber\\
             &&-\frac{1}{8}\,\lm^2{\x}^{5}-\frac{1}{16}\,{\lm}^{4}\x^4+{\frac{99}{128}}\,{\lm}^{4}\x^5,\\
 t^{(2)}_{0,2} &=& -\frac{1}{8}\,{\x}^{2}-\frac{1}{8}\,{\x}^{3}-{\frac{5}{64}}\,{\x}^{4}-{\frac{5}{512}}\,
{\x}^{5}+\frac{1}{2}\,\lm^2\x^3\\
            &&+{\frac{59}{384}}\,{\lm}^{2}{\x}^{5}+{\frac{21}{32
}}\,{\lm}^{2}{\x}^{4}+\frac{3}{16}\,{\lm}^{4}\x^4-{\frac{53}{64}}\,{\lm}^{4}\x^5,\nonumber
\\
 t^{(2)}_{1,1} &=& -{\frac{5}{64}}\,{\x}^{4}-{\frac{45}{256}}\,{\x}^{5}-{\frac{137}{192}
}\,{\lm}^{2}{\x}^{5}-{\frac{5}{16}}\,{\lm}^{2}{\x}^{4}\nonumber\\
             &&+\frac{1}{8}\,{\lm}^{4}\x^4-{\frac{9}{16}}\,{\lm}^{4}\x^5\\
 t^{(3)}_{0,1} &=& {\frac{3}{1024}}\,{\x}^{5}-{\frac{7}{384}}\,{\lm}^{2}{\x}^{5}-{\frac{7}
{64}}\,{\lm}^{4}\x^5,\\
 t^{(3)}_{0,2} &=& \frac{1}{32}\,{\x}^{4}+{\frac{57}{1024}}\,{\x}^{5}+{\frac{13}{768}}\,{\lm}^{2}{\x
}^{5}-{\frac{31}{128}}\,{\lm}^{4}\x^5\\
 t^{(3)}_{0,3} &=& \frac{1}{16}\,{\x}^{3}+\frac{1}{16}\,{\x}^{4}-{\frac{5}{512}}\,{\x}^{5}+\frac{3}{16}\,{\lm}^{2}{\x}^{4}+{\frac{19}{24}}\,{\lm}^{2}{\x}^{5}\nonumber\\
             &&+{\frac{33}{64}}\,{\lm}^{4}\x^5,
\end{eqnarray}
\begin{eqnarray}
 t^{(3)}_{1,2} &=& {\frac{7}{512}}\,{\x}^{5}-{\frac{1}{96}}\,{\lm}^{2}{\x}^{5}+\frac{3}{16}\,{\lm}^{4
}\x^5,\\
 t^{(4)}_{0,0} &=& {\frac{35}{64}}\,{\x}^{4}+{\frac{41}{64}}\,{\x}^{5}-{\lm}^{2}{\x}^{4}-\frac{5}{8}
\,{\lm}^{2}{\x}^{5}-\frac{1}{8}\,{\lm}^{4}\x^5,\\
 t^{(4)}_{0,1} &=& -\frac{1}{16}\,{\x}^{4}-{\frac{23}{512}}\,{\x}^{5}+{\frac{5}{16}}\,{\lm}^{2}{\x}^
{4}-{\frac{11}{192}}\,{\lm}^{2}{\x}^{5}\nonumber\\
             &&-{\frac{9}{32}}\,{\lm}^{4}\x^5,\\
 t^{(4)}_{0,2} &=& -{\frac{3}{64}}\,{\x}^{5}+\frac{3}{8}\,{\lm}^{2}{\x}^{4}+{\frac{13}{16}}\,{\lm}^{2
}{\x}^{5}+{\frac{11}{16}}\,{\lm}^{4}\x^5,\\
 t^{(5)}_{0,1} &=& {\frac{1}{256}}\,{\x}^{5}+\frac{1}{32}\,{\lm}^{2}{\x}^{5}+{\frac{7}{64}}\,{\lm}^{4
}\x^5,\\
 t^{(5)}_{0,2} &=& {\frac{1}{128}}\,{\x}^{4}+{\frac{35}{512}}\,{\x}^{5}-\frac{3}{16}\,{\lm}^{2}{\x}^
{4}-{\frac {325}{384}}\,{\lm}^{2}{\x}^{5}\nonumber\\
            &&-{\frac{7}{64}}\,{\lm}^{4}\x^5, \\
 t^{(6)}_{0,3} &=& -{\frac{19}{1024}}\,{\x}^{5}-{\frac{5}{384}}\,{\lm}^{2}{\x}^{5},\\
 t^{(6)}_{0,4} &=& -{\frac{5}{128}}\,{\x}^{4}-{\frac{3}{64}}\,{\x}^{5}+\frac{1}{16}\,{\lm}^{2}{\x}^{
5},
\\
 t^{(7)}_{0,4} &=& -{\frac{9}{64}}\,{\x}^{5}-{\frac{13}{16}}\,{\lm}^{2}{\x}^{5},\\
 t^{(7)}_{1,4} &=& -{\frac{3}{128}}\,{\x}^{5}+{\frac{7}{32}}\,{\lm}^{2}{\x}^{5},\\
 t^{(7)}_{2,4} &=& \frac{3}{8}\,{\lm}^{2}{\x}^{5},\\
 t^{(10)}_{0,5} &=& \frac{7}{256} \x^5,\\
 t^{(11)}_{0,4} &=& -{\frac{5}{1024}}\,{\x}^{5}-{\frac{7}{384}}\,{\lm}^{2}{\x}^{5}.
\end{eqnarray}

Since the Hamiltonian is hermitian, the hopping amplitudes obey $t^{(n)}_{i,j}=t^{(n)}_{j,i}$ which allows to reduce the numerical effort.

For clarity, we also give the embedding factors for each hopping element. Let $\nu^{(n)}_{i,j}$ denote the embedding factor of graph $\mathcal{G}_n$ associated with the hopping element $t^{(n)}_{i,j}$. Note that here we have omitted the obvious dependence on the spatial dimension $d$. These factors can be interpreted as the embedding factors of graph $\mathcal{G}_n$ when the part of the graph connecting site $i$ and site $j$ is already embedded. In general, this leads to a case distinction. The embedding factors by construction obey $\nu^{(n)}_{i,j}=\nu^{(n)}_{j,i}$ and are given by
\begin{alignat}{2}
 \nu^{(0)}_{0,0} &= 1, \\
 \nu^{(1)}_{0,0} &= 2 d,      &&\nu^{(1)}_{0,1} = 1, \notag \\
 \nu^{(2)}_{0,0} &= 2d(2d-1), &&\nu^{(2)}_{0,1} = 2d-1, \notag \\
 \nu^{(2)}_{0,2} &= 1,        &&\nu^{(2)}_{1,1} = d(2d-1), \notag\\
 \nu^{(3)}_{0,1} &= (2d-1)(2d-1),~~&& \nu^{(3)}_{0,2} = 2d-1, \notag \\
 \nu^{(3)}_{0,3} &= 1, &&\nu^{(3)}_{1,2} = (2d-1)(2d-1), \notag
\end{alignat}
\begin{alignat}{2}
 \nu^{(4)}_{0,0} &= 2d(d-1), &&\nu^{(4)}_{0,1} = 2(d-1), \notag\\
 \nu^{(4)}_{0,2} &= 1, &&\nu^{(5)}_{0,1} = (2d-1)(d-1), \notag\\
 \nu^{(5)}_{0,2} &= 2d-2, &&\nu^{(6)}_{0,3} = 2d-1, \notag\\
 \tilde{\nu}^{(6)}_{0,3} &= 2d-2, &&\nu^{(6)}_{0,4} = 1, \notag
 \\
 \nu^{(7)}_{0,4} &= 2(d-1)+(d-1)2(d&&-2), \notag
\end{alignat}
\begin{alignat}{2}
 \nu^{(7)}_{1,4} &= 2(d-1), &&\tilde{\nu}^{(7)}_{1,4} = 2(d-1)-1, \notag\\
 \nu^{(7)}_{2,4} &= 1, \notag\\
 \nu^{(10)}_{0,5} &= 1, &&\!\!\nu^{(11)}_{0,4} = 2(d-1). \notag
\end{alignat}
\begin{widetext}
Finally, we give the specific contributions to the one-triplon dispersion after Fourier transformation. Let $\omega_n(\vec{k})$ denote the contribution of graph $\mathcal{G}_n$ to the one-particle dispersion $\Wk$. The contributions of the graphs read
\begin{eqnarray}
 \omega_0 &=& \nu^{(0)}_{0,0} t^{(0)}_{0,0}\\
 \omega_1 &=& \nu^{(1)}_{0,0} t^{(1)}_{0,0}+ \sum_{\bar{k}\in \{\pm k_1,\ldots,\pm k_d \}} \nu^{(1)}_{0,1} t^{(1)}_{0,1} \cos (\bar{k})\\
 \omega_2 &=& \nu^{(2)}_{0,0} t^{(2)}_{0,0}+\nu^{(2)}_{1,1} t^{(2)}_{1,1}+ \sum_{\bar{k}} \nu^{(2)}_{0,1} t^{(2)}_{0,1} \cos (\bar{k})+\sum_{\bar{k}_1}  \sum_{\bar{k}_2 \neq -\bar{k}_1} \nu^{(2)}_{0,2} t^{(2)}_{0,2} \cos (\bar{k}_1+\bar{k}_2)\\
 \omega_3 &=& \sum_{\bar{k}} (2\nu^{(3)}_{0,1} t^{(3)}_{0,1}+\nu^{(3)}_{1,2} t^{(3)}_{1,2}) \cos (\bar{k})+\sum_{\bar{k}_2 \neq -\bar{k}_1} 2 \nu^{(3)}_{0,2} t^{(3)}_{0,2} \cos (\bar{k}_1+\bar{k}_2)\\
          &+& \sum_{\bar{k}_1}  \sum_{\bar{k}_2 \neq -\bar{k}_1} \sum_{\bar{k}_3 \neq -\bar{k}_2} \nu^{(3)}_{0,3} t^{(3)}_{0,3} \cos (\bar{k}_1+\bar{k}_2+\bar{k}_3)\\
 \omega_4 &=& \nu^{(4)}_{0,0} t^{(4)}_{0,0}+ \sum_{\bar{k}} \nu^{(4)}_{0,1} t^{(4)}_{0,1} \cos (\bar{k}) +\tfrac{1}{2} \sum_{\bar{k}_1}  \sum_{\bar{k}_2 \neq \pm \bar{k}_1} \nu^{(4)}_{0,2} t^{(4)}_{0,2} \cos (\bar{k}_1+\bar{k}_2)\\
 \omega_5 &=& \sum_{\bar{k}} 2 \nu^{(5)}_{0,1} t^{(5)}_{0,1} \cos (\bar{k}) +\sum_{\bar{k}_1}  \sum_{\bar{k}_2 \neq -\bar{k}_1} \nu^{(5)}_{0,2} t^{(5)}_{0,2} \cos (\bar{k}_1+\bar{k}_2)\\
 \omega_6 &=& \sum_{\bar{k}_1}  \sum_{\bar{k}_2 \neq -\bar{k}_1} \sum_{\bar{k}_3 \neq -\bar{k}_2}  \sum_{\bar{k}_4 \neq -\bar{k}_3 \atop \bar{k}_1+\bar{k}_2+\bar{k}_3+\bar{k}_4\neq 0} \nu^{(6)}_{0,4} t^{(6)}_{0,4} \cos (\bar{k}_1+\bar{k}_2+\bar{k}_3+\bar{k}_4) \quad\\
          &+& \sum_{\bar{k}_1}  \sum_{\bar{k}_2 \neq -\bar{k}_1} \sum_{\bar{k}_3 \neq -\bar{k}_2} \begin{cases}
  2 \tilde{\nu}^{(6)}_{0,3} t^{(6)}_{0,3} \cos (\bar{k}_1+\bar{k}_2+\bar{k}_3),  & \text{if } \bar{k}_1=-\bar{k}_3 \text{ and }\bar{k}_1\neq \bar{k}_2 \\
  2 \nu^{(6)}_{0,3} t^{(6)}_{0,3} \cos (\bar{k}_1+\bar{k}_2+\bar{k}_3), & \text{else}
\end{cases} \\
 \omega_7 &=& \sum_{\bar{k}} 2 \nu^{(7)}_{0,4} t^{(7)}_{0,4} \cos (\bar{k}) \\
          &+& \frac{1}{2} \sum_{\bar{k}_1}  \sum_{\bar{k}_2 \neq \pm \bar{k}_1} \sum_{\bar{k}_3 \neq -\bar{k}_2 \atop \bar{k}_3 \neq -\bar{k}_1} 2 \nu^{(7)}_{2,4} t^{(7)}_{2,4} \cos (\bar{k}_1+\bar{k}_2+\bar{k}_3)\\
          &+& \sum_{\bar{k}_1}  \sum_{\bar{k}_2 \neq -\bar{k}_1} \begin{cases}
  2 \tilde{\nu}^{(7)}_{1,4} t^{(7)}_{1,4} \cos (\bar{k}_1+\bar{k}_2),  & \text{if } \bar{k}_1\neq \bar{k}_2 \\
  2 \nu^{(7)}_{1,4} t^{(7)}_{1,4} \cos (\bar{k}_1+\bar{k}_2), & \text{else}
\end{cases} \\
 \omega_{10} &=& \sum_{\bar{k}_1}  \sum_{\bar{k}_2 \neq -\bar{k}_1} \sum_{\bar{k}_3 \neq -\bar{k}_2}  \sum_{\bar{k}_4 \neq -\bar{k}_3 \atop \bar{k}_1+\bar{k}_2+\bar{k}_3+\bar{k}_4\neq 0} \sum_{\bar{k}_5 \neq -\bar{k}_4 \atop \bar{k}_2+\bar{k}_3+\bar{k}_4+\bar{k}_5\neq 0} \nu^{(10)}_{0,5} t^{(10)}_{0,5} \cos (\bar{k}_1+\bar{k}_2+\bar{k}_3+\bar{k}_4+\bar{k}_5)\\
 \omega_{11} &=& \sum_{\bar{k}_1}  \sum_{\bar{k}_2 \neq -\bar{k}_1} \sum_{\bar{k}_3 \neq -\bar{k}_2} 2 \nu^{(11)}_{0,4} t^{(11)}_{0,4} \cos (\bar{k}_1+\bar{k}_2+\bar{k}_3) \quad.
\end{eqnarray}
\end{widetext}
The one-triplon dispersion is just the sum of these contributions, $\Wk=\sum_n \omega_n(\vec{k})$. It is convenient to convert the restricted momentum summations over cosines into combinations
of $\gk$, $\gamma_{2\vec k}$ etc. This can be done in a straightforward fashion using trigonometric theorems, yielding $\sum_{\bar{k}}\cos \bar{k} = 2d\gk$,
$\sum_{\bar{k}_1}  \sum_{\bar{k}_2 \neq -\bar{k}_1} \cos (\bar{k}_1+\bar{k}_2) = 4d^2 \gk^2 - 2d$ etc.



\begin{thebibliography}{99}
\bibitem{scalenote} Taking the limit $d\to\infty$ at fixed $\qq=Kd/J$ ensures that both intra-dimer and inter-dimer interactions yield contributions of order $d^0$ to the energy per site. If one would instead take the limit $d\to\infty$ at fixed $K/J$ in \eqref{hh}, the large number of neighbors in $\sum_{\langle ii'\rangle}$ would simply suppress the effect of $J$ such that the ground state of \eqref{hh} would be that of two decoupled $d$-dimensional cubic antiferromagnets.
\bibitem{anomfoot}
The anomalous expectation value $Q(d) = \sum_i \langle t_{i\alpha}^\dagger t_{i\alpha}^\dagger \rangle/N$ must vanish for any physical eigenstate of $\mathcal{H}$ in any dimension $d$. The $1/d$ expansion calculates $\bar Q(x) = Q(1/x)$ in a Taylor series in $x$. According to our explicit results, the first two coefficients of this expansion vanish. Together with $\bar Q(x)=0$ for any non-zero $x$ this implies that $\bar Q(x)=0$ order by order.
\end{thebibliography}
\end{document}